\RequirePackage{ifpdf}
\ifpdf % We are running pdfTeX in pdf mode
\documentclass[pdftex]{sigma}
\else
\documentclass{sigma}
\fi

\def\dbigoplus{\mathop{\displaystyle \bigoplus }}
\def\func#1{\mathop{\rm #1}\nolimits}

\begin{document}

\allowdisplaybreaks

\renewcommand{\PaperNumber}{043}

\FirstPageHeading

\ShortArticleName{Supersymmetry Flows of Af\/f\/ine Toda Models}

\ArticleName{Supersymmetry of Af\/f\/ine Toda Models as Fermionic\\
Symmetry Flows of the Extended mKdV Hierarchy}

\Author{David M. SCHMIDTT}

\AuthorNameForHeading{D.M. Schmidtt}

\Address{Instituto de F\'{\i}sica Te\'{o}rica, UNESP-Universidade Estadual Paulista,\\
Caixa Postal 70532-2, 01156-970, S\~{a}o Paulo, SP, Brasil}

\Email{\href{mailto:david@ift.unesp.br}{david@ift.unesp.br}, \href{mailto:david.schmidtt@gmail.com}{david.schmidtt@gmail.com}}

\ArticleDates{Received December 10, 2009, in f\/inal form May 19, 2010;  Published online May 27, 2010}

\Abstract{We couple two copies of the supersymmetric mKdV hierarchy
by means of the algebraic dressing technique. This allows to deduce
the whole set of $(N,N)$ supersymmetry transformations of the
relativistic sector of the extended mKdV hierarchy and to interpret
them as fermionic symmetry f\/lows. The construction is based on an
extended Riemann--Hilbert problem for af\/f\/ine Kac--Moody superalgebras
with a half-integer gradation. A~gene\-ra\-lized set of
relativistic-like fermionic local current identities is introduced
and it is
shown that the simplest one, corresponding to the lowest isospectral times $t_{\pm 1}$ provides the supercharges generating rigid supersymmetry
transformations in 2D superspace. The number of supercharges is
equal to the dimension of the fermionic kernel of a given semisimple
element $E\in \widehat{\mathfrak{g}}$ which def\/ines both, the
physical degrees of freedom
and the symmetries of the model. The general construction is applied to the $N=(1,1)$ and $N=(2,2)$ sinh-Gordon models which are worked out in
detail.}


\Classification{81T60; 37K20; 37K10}

\section{Introduction}

It is well known that bosonic Toda models are underlined by Lie
algebras and that they provide some sort of f\/ield theoretic
realization to them. They are relevant to particle physics because
they describe integrable perturbations of two-dimensional conformal
f\/ield theories, allow soliton conf\/igurations in their spectrum and
are useful laboratories to develop new methods relevant to the study
of non-perturbative aspects of quantum f\/ield theory.

A natural step when having a bosonic f\/ield theory is to try to
incorporate fermions and to construct its supersymmetry extension.
In the case of bosonic Toda models this is a not an easy task
because we want to preserve the integrability, which is one of the
main properties of this kind of theories. Integrability is a
consequence of the existence of an inf\/inite number of bosonic
Hamiltonians in involution which depend strongly on the Lie
algebraic input data def\/ining the Toda model itself. Each
Hamiltonian generates a bosonic (even) symmetry f\/low and due to the
fact that supersymmetry is just a symmetry, it is natural to expect
the presence of conserved supercharges each one generating its own
fermionic (odd) symmetry f\/low and also to expect that the
supersymmetric extension is not related to a Lie algebra but to a
Lie superalgebra, see \cite{Evans-Madsen} for an example of how
bosonic symmetries are not preserved after supersymmetrization. By
def\/inition, a supersymmetry is a symmetry where the application of
two successive odd transformations close into an even one. If there
is an inf\/inite number of even f\/lows, then it is natural to
incorporate the same number of odd f\/lows in order to close the `f\/low
superalgebra'. Hence, the set of f\/ields $\mathcal{F}$ will depend on
an inf\/inite number of even and odd variables $\mathcal{F}=\mathcal{F}(t_{\pm
1/2},t_{\pm 1},t_{\pm 3/2},t_{\pm 3},\dots)$, see
\cite{manin-radul} for a f\/irst example of this `f\/low approach'
applied to the KP hierarchy. Our main motivation to formulate
supersymmetric af\/f\/ine Toda models within this setting relies on the
possibility of using powerful techniques available in the theory of
inf\/inite-dimensional Lie algebras and integrable systems, in
particular, vertex operator representations and tau functions. The
goal is to set the ground to study the quantization of the af\/f\/ine
super Toda integrable models within this fashion.

Several authors have studied the problem of constructing
supersymmetric extensions of integrable hierarchies. On one side,
for the Toda lattice most of them use superf\/ields as a natural way
to supersymmetrize Lax operators while preserving integrability or
to obtain a manifestly supersymmetric
Hamiltonian reduction of super WZNW models, see for example \cite{Olshanetsky,Evans,Sorba,Spence}. The common
conclusion is that only Lie superalgebras (classical or af\/f\/ine) with
a purely fermionic simple root system allow supersymmetric
integrable extensions, otherwise supersymmetry is broken. On the
other side, there are several supersymmetric formulations of the
Drinfeld--Sokolov reduction method for constructing integrable
hierarchies in which the algebraic Dressing method and the `f\/low
approach' were gradually developed and worked out in several
examples, see
for example \cite{Inami-Kanno,Delduc-Gallot,Miramontes01,symmetry flows}. The main goal of these works is the construction
of an inf\/inite set of fermionic non-local symmetry f\/lows but a clear
relation between the conserved supercharges and its corresponding
f\/ield component transformations remains obscure. In
\cite{half-integer gradations}, fermionic f\/ields were coupled to the
Toda f\/ields in a supersymmetric way in the spirit of generalized
Toda models coupled to matter f\/ields introduced in
\cite{Gervais-Saveliev} and further analyzed in~\cite{Ferreira}.
This coupling was performed on-shell and only the f\/irst half of the
supersymmetric sector was analyzed (corresponding to the positive part $t_{+1/2}$, $t_{+1}$). An important result of that paper was the
introduction
of a `reductive' automorphism $\tau _{\rm red}$ (constructed explicitly in the $sl(2,1)$ af\/f\/ine case) devised to remove the non-locality of the
lowest supersymmetric f\/low $t_{+1/2}$, as a consequence, it was
shown that it is not strictly necessary to start with an af\/f\/ine
superalgebra with a purely fermionic simple root system in order to
get an integrable supersymmetric extension of a bosonic model. See
also~\cite{toppan} for another (based on Osp(1,4) having one bosonic
and one fermionic simple roots) example of a~Toda model with
superconformal symmetry realized non-linearly. The complementary
of\/f-shell Hamiltonian reduction was developed in~\cite{yo} by using
a two-loop super-WZNW model where the (local) action functional
leading to the supersymmetric Leznov--Saveliev equations of motion
was constructed, in principle, for any superalgebra endowed with a
half-integer gradation and invariant under $\tau _{\rm red}$. It was
also shown that several known purely fermionic integrable models
belong to the family of perturbed WZNW on supercosets where the
bosonic part is fully gauged away.

The purpose of this paper is to introduce the second half of the
supersymmetric sector (corresponding to the negative part
$t_{-1/2},t_{-1}$) and to study the whole coupled system generated
by the subset of symmetry f\/lows $(t_{-1},t_{-1/2},t_{+1/2},t_{+1})$.
This analysis was not performed neither in \cite{half-integer
gradations} nor \cite{yo} so this work complement their study. The
outcome is that the supersymmetry f\/lows described in terms of the
algebraic dressing technique turn out to be equivalent to the usual
notion of supersymmetry described in terms of superspace variables
(this is shown by considering explicit examples). This allows to
locate the supersymmetry of the models inside a~formalism which is
manifestly integrable by construction.

In Section~\ref{section2.1} we review the algebraic dressing technique and use it
to couple two identical copies of the same integrable hierarchy thus
def\/ining its extension. In Section~\ref{section2.2} we introduce the
relativistic/supersymmetric sector of the extended super-mKdV
hierarchy by coupling two super-mKdV
hierarchies in dif\/ferent gauges. This idea was f\/irst used in \cite{Miramontes02} in the bosonic case. In Section~\ref{section2.3} we construct two
odd Lax pairs associated to the chiral sectors of the hierarchy and
in Section~\ref{section2.4} the complete set of extended $(N,N)$ supersymmetry
transformations is given. The recursion operators are given in
Section~\ref{section2.5} to show that all higher fermionic f\/lows are non-local.
In Section~\ref{section2.6} we use the extended Riemann--Hilbert problem to
construct a~set of local fermionic current identities associated to
the non-Abelian f\/lows $t_{\pm 1/2}$, where each pair of isospectral
f\/lows $t_{\pm n}$ is coupled in a relativistic-like manner. It is
also shown that the number of supersymmetries (i.e.\ supercharges) is
equal to the dimension of the fermionic kernel of the
operator ${\rm Ad}\, E$. This means that in the superalgebra decomposition $\widehat{\mathfrak{g}} =\ker \left( {\rm Ad}\, E\right) \oplus
\func{Im}\left( {\rm Ad}\,E\right) $ induced by the constant semisimple
element $E$, any element of the af\/f\/ine superalgebra~$\widehat{\mathfrak{g}}$ have a well def\/ined role, i.e.\ it def\/ines a
symmetry f\/low or a physical degree of freedom. In Section~\ref{section2.7} we use
the two-Loop super WZNW action to construct the supercharges
generating the supersymmetry transformations giving a direct
relation between the $t_{\pm 1/2}$ odd f\/lows and the f\/ields
transformations. We also show that the Noether procedure reproduces
the supercharges constructed in~Section~\ref{section2.6} by using the
factorization problem thus conf\/irming their equivalence. Finally, in
the Sections~\ref{section3.1} and~\ref{section3.2} we study in detail the construction in
order to have a better feeling of how the fermionic symmetry
f\/lows of the models are def\/ined by the kernel part ($\ker \left( {\rm Ad}\,E\right))$ and to make contact with the usual notion of superspace. We also
give an example of a solution to a relativistic-like equation
expressed in terms of the higher graded $t_{\pm 3}$ isospectral
times only, thus generalizing the sine-Gordon equation. In the
conclusion we pose the more important problems to be treated in the
future which are the main motivations of the present work.

\section{General analysis}\label{section2}

Here we study the supersymmetric sector of the extended mKdV
hierarchy and obtain the main results of the paper. The goal of this
chapter is to put into one consistent body the new pieces with the known previous results. The core of the f\/low approach we will follow relies on
the algebraic dressing technique used to unify symmetry f\/lows
(isospectral and non-Abelian) of integrable hierarchies related to
af\/f\/ine Lie algebras. The Riemann--Hilbert factorization def\/ines the
integrable structure and a related hierarchy of non-linear partial
dif\/ferential equations.

\subsection{The algebraic dressing technique}\label{section2.1}

Consider an af\/f\/ine Lie superalgebra
$\widehat{\mathfrak{g}}=\underset{i\in
\mathbb{Z}
/2=-\infty }{\overset{+\infty }{\dbigoplus
}}\widehat{\mathfrak{g}}_{i}$
half-integer graded by an operator $Q$ ($\left[ Q,\widehat{\mathfrak{g}}_{i}%
\right] =i\widehat{\mathfrak{g}}_{i}$) and two supergroup elements
(dressing matrices) $\Theta $ and $\Pi $ taken as the exponentials
of the
negative/positive subalgebras of $\widehat{\mathfrak{g}}$ respectively, i.e.\ $\widehat{\mathfrak{g}}_{-}$ and $\widehat{\mathfrak{g}}_{+}$ in the
decomposition $\widehat{\mathfrak{g}}=\widehat{\mathfrak{g}}_{-}+\widehat{%
\mathfrak{g}}_{+}$ induced by the projections $\mathcal{P}_{\pm
}(\ast )=(\ast )_{\pm }$ along positive and negative grades. They
are taken to be
formal expansions of the form%
\begin{gather}
\Theta  = \exp \big( \psi ^{\left( -1/2\right) }+\psi ^{\left(
-1\right)
}+\psi ^{\left( -3/2\right) }+\cdots\big), \notag  \\
\Pi =BM,\qquad M=\exp \big( -\psi ^{\left( +1/2\right) }-\psi
^{\left( +1\right) }-\psi ^{\left( +3/2\right) }-\cdots\big),\label{Dressing matrices}
\end{gather}%
where $B=\exp \widehat{\mathfrak{g}}_{0}\in \widehat{G}_{0}$ and
$\psi ^{\left( i\right) }\in \widehat{\mathfrak{g}}_{i}$. The
constant semisimple elements $E^{(\pm 1)}$ of grade~$\pm 1$ $\left(
\left[ Q,E^{(+1)}\right] =\pm E^{(+1)}\right) $ def\/ine operators
${\rm Ad}\,E^{(\pm 1)}$ each one splitting the superalgebra
$\widehat{\mathfrak{g}}=\mathcal{K}^{\pm }+\mathcal{M}^{\pm
}$ into kernel and image subspaces obeying $\left[ \mathcal{K}^{\pm },%
\mathcal{K}^{\pm }\right] \subset \mathcal{K}^{\pm }$, $\left[ \mathcal{K}%
^{\pm },\mathcal{M}^{\pm }\right] \subset \mathcal{M}^{\pm }$, where $%
\mathcal{K}^{\pm }\equiv \ker \left( {\rm Ad}\, E^{(\pm 1)}\right) $ and $\mathcal{M}%
^{\pm }\equiv \func{Im}\left( {\rm Ad}\,E^{(\pm 1)}\right) $. The kernel and
image
subspaces have bosonic and fermionic components $\mathcal{K}^{\pm }=\mathcal{%
K}_{B}^{\pm }\oplus \mathcal{K}_{F}^{\pm }$ and $\mathcal{M}^{\pm }=\mathcal{%
M}_{B}^{\pm }\oplus \mathcal{M}_{F}^{\pm }$ each one having a well
def\/ined (half-integer) grade respect the operator $Q$.

Recall \cite{babelon-bernad} that the dressing transformation of
$x\in \widehat{G}$ by $g\in \widehat{G}$ is def\/ined by
\[
{}^{g}x=\left(
xgx^{-1}\right) _{\pm }xg_{\pm }^{-1}.
\] The inf\/initesimal
transformation for
$g=\exp A$ with $A=A_{+}+A_{-}$ $\ $and $A_{\pm }\in \widehat{\mathfrak{g}}%
_{\pm }$ is
\begin{equation}
\delta _{A}x= {}^{g}x-x=\pm \left( xAx^{-1}\right) _{\pm }x\mp xA_{\pm
}. \label{Infinitesimal action}
\end{equation}%
From this we f\/ind the pure actions of $A=A_{+}\in \mathcal{K}_{+}$ and $%
A=A_{-}\in \mathcal{K}_{-}$ on $x=\Theta $ and $x=\Pi $, respectively%
\begin{equation}
\delta _{A_{+}}\Theta =-\left( \Theta A_{+}\Theta ^{-1}\right)
_{-}\Theta,\qquad \delta _{A_{-}}\Pi =+\left( \Pi A_{-}\Pi
^{-1}\right) _{+}\Pi.  \label{positive variarion}
\end{equation}%
To see this, consider $A=A_{+}$ and $x=\Theta $ and the upper sign in (\ref{Infinitesimal action}). We get
\[
\delta _{A_{+}}\Theta =\big(
\Theta A_{+}\Theta ^{-1}\big) _{+}\Theta -\Theta A_{+}=-\big(
\Theta A_{+}\Theta ^{-1}\big) _{-}\Theta,
\]
 where we have used the
decomposition $\Theta A_{+}\Theta ^{-1}=\left( \Theta A_{+}\Theta
^{-1}\right) _{+}+\left( \Theta A_{+}\Theta ^{-1}\right) _{-}$. For
$A=A_{-}$ and $x=\Pi $ the proof is similar. We also have that for
$A=A_{-}$ and $x=\Theta $ and for $A=A_{+}$ and $x=\Pi $ the
variations vanish, $\delta _{A_{-}}\Theta =0$ and $\delta
_{A_{+}}\Pi =0$ respectively. Hence, in the present form, the
dressing matrices~(\ref{Dressing matrices}) only evolve under half
of the f\/lows.

Setting $A_{\pm }=$ $t_{\pm n}E^{(\pm n)}$ $($with $\left[ Q,E^{(\pm n)}%
\right] =\pm nE^{(\pm n)})$ and taking the limit $t_{\pm
n}\rightarrow 0,$
we have the isospectral evolutions for $\Theta $ and $\Pi $%
\begin{equation}
\partial _{+n}\Theta =-\big( \Theta E^{(+n)}\Theta ^{-1}\big) _{-}\Theta
,\qquad \partial _{-n}\Pi =+\big( \Pi E^{(-n)}\Pi
^{-1}\big) _{+}\Pi,  \label{Isospectral evolutions}
\end{equation}%
where $\delta _{A_{+}}\Theta /t_{+}=\left( ^{A_{+}}\Theta -\Theta
\right) /t_{+}\rightarrow \partial _{+n}\Theta $ and similar for
$\delta _{A_{-}}\Pi $. From equations (\ref{Isospectral evolutions})
we obtain the dressing relations
\begin{gather*}
E_{\Theta }^{(+n)}=\big( \Theta
E^{(+n)}\Theta ^{-1}\big) _{+}=\Theta E^{(+n)}\Theta
^{-1}+\partial _{+n}\Theta \Theta ^{-1},\\
E_{\Pi }^{(-n)}=\big(
\Pi E^{(-n)}\Pi ^{-1}\big) _{-}=\Pi E^{(-n)}\Pi ^{-1}-\partial
_{-n}\Pi \Pi ^{-1}
\end{gather*}
 and the Lax operators
 \[
 L_{+n}=\partial
_{+n}-E_{\Theta }^{(+n)},\qquad L_{-n}=\partial _{-n}+E_{\Pi }^{(-n)}.
\]
The
Baker--Akhiezer wave functions $\Psi _{\pm }$ are def\/ined by $L_{\pm n}\Psi _{\mp }$ $%
=0$ and are given by
\begin{equation*}
\Psi _{-}=\Theta \exp \left( +\sum\limits_{n\in
\mathbb{Z}
^{+}}t_{+n}E^{(+n)}\right),\qquad \Psi _{+}=\Pi \exp \left(
-\sum\limits_{n\in
\mathbb{Z}
^{+}}t_{-n}E^{(-n)}\right).
\end{equation*}

Equations (\ref{Isospectral evolutions}) describe two identical but
decoupled systems of evolution equations as shown above, the
coupling of the two sectors is achieved by imposing the relation
$g=\Psi _{-}^{-1}\Psi _{+}$
with $g$ a constant group element. Alternatively, we have
\begin{equation}
\exp \left( +\sum\limits_{n\in
\mathbb{Z}
^{+}}t_{+n}E^{(+n)}\right) g\exp \left( +\sum\limits_{n\in
\mathbb{Z}
^{+}}t_{-n}E^{(-n)}\right) =\Theta ^{-1}(t)\Pi (t).  \label{RH}
\end{equation}%
This is the extended Riemann--Hilbert factorization problem
originally used in \cite{top-antitop fusion} to extend the mKdV
hierarchy to the negative f\/lows. From (\ref{RH}) we recover
(\ref{Isospectral evolutions}) and two important extra equations
describing the isospectral evolution of $\Theta $ and $\Pi $ with
respect opposite f\/low parameters
\begin{gather}
\partial _{+n}\Pi =+\big( \Theta E^{(+n)}\Theta ^{-1}\big) _{+}\Pi,\qquad \partial _{-n}\Theta =-\big( \Pi E^{(-n)}\Pi ^{-1}\big)
_{-}\Theta.  \label{Isospectral II}
\end{gather}%
These equations are extended to actions of $A_{+}\in \mathcal{K}_{+}$ and $%
A_{-}\in \mathcal{K}_{-}$ on $\Pi $ and\ $\Theta $, similar to (\ref{positive variarion}) we have%
\begin{equation}
\delta _{A_{+}}\Pi =+\big( \Theta A_{+}\Theta ^{-1}\big) _{+}\Pi
,\qquad \delta _{A_{-}}\Theta =-\big( \Pi A_{-}\Pi
^{-1}\big) _{-}\Theta.  \label{negative variation}
\end{equation}

The equations (\ref{positive variarion}), (\ref{Isospectral
evolutions}) and (\ref{Isospectral II}), (\ref{negative variation})
describe the isospectral
evolution and non-Abelian variations of the dressing matrices $\Theta $ and $%
\Pi $ and their consistency, as an algebra of f\/lows, is encoded in
Proposition~\ref{proposition1} below. Note that the f\/lows associated to the
positive times
are dual to the ones associated to the negative times, in the sense that $%
\mathcal{K}_{+}^{\ast }\simeq \mathcal{K}_{-}$ under the (assumed to
exists) non-degenerate inner product $\left\langle \ast
\right\rangle $ which provide the orthogonality condition $\left\langle \widehat{\mathfrak{g}}_{i}\widehat{%
\mathfrak{g}}_{j}\right\rangle =\delta _{i+j,0}$ of
graded spaces. This also show how
the degrees of freedom are naturally doubled by the extension.

\begin{remark}
If we consider pseudo-dif\/ferential operators, the equations (\ref{Isospectral evolutions}), (\ref{Isospectral II}) are good starting
points to extend the KP hierarchy with the negative f\/lows and the
expectation value of (\ref{RH}) to extend its corresponding $\tau$-function.
\end{remark}

\subsection{Relativistic sector of the extended mKdV hierarchy}\label{section2.2}

From (\ref{Isospectral evolutions}) and (\ref{Isospectral II}) we
have the following

\begin{definition}
The relativistic sector of the extended mKdV hierarchy is def\/ined by
the
following set of evolution equations%
\begin{alignat}{3}
& \partial _{+}\Theta =+\big( \Theta E_{+}^{(+1)}\Theta ^{-1}\big)
_{<0}\Theta,\qquad && \partial _{+}\Pi =-\big( \Theta
E_{+}^{(+1)}\Theta ^{-1}\big) _{\geq 0}\Pi,  & \notag
 \\
& \partial _{-}\Theta  = +\big( \Pi E_{-}^{(-1)}\Pi ^{-1}\big) _{<0}\Theta
,\qquad && \partial _{-}\Pi =-\big( \Pi E_{-}^{(-1)}\Pi
^{-1}\big) _{\geq 0}\Pi,& \label{isospectral evolution}
\end{alignat}%
for the two isospectral times $t_{\pm 1}=-x^{\pm }$ associated to the grade $%
\pm 1$ constant elements $E_{\pm }^{(\pm 1)}\in
\widehat{\mathfrak{g}}.$ The $\left( \ast \right) _{\geq 0}$ denote
projection onto grades $\geq 0$ and the $\left( \ast \right) _{<0}$
onto grades $\leq -1/2.$
\end{definition}

In the def\/inition above we write explicitly the projections $\left(
\ast \right) _{\pm }$ in terms of grades in order to avoid confusion
with the dif\/ferent projections used below in (\ref{Gauge isospectral
evolution}). The
Lax covariant derivative $\left( L=d+A^{L}\right) $ extracted from (\ref{isospectral evolution}) has a Lax connection $A^{L}$ given by
\begin{alignat}{3}
& L_{-}  = \partial _{-}+A_{-}^{L},\qquad && A_{-}^{L}=-B\big(
E_{-}^{(-1)}+\psi _{-}^{\left( -1/2\right) }\big) B^{-1},& \notag
\\
& L_{+} =\partial _{+}+A_{+}^{L},\qquad && A_{+}^{L}=-\partial
_{+}BB^{-1}+\psi _{+}^{\left( +1/2\right) }+E_{+}^{(+1)}, & \label{Lax pair}
\end{alignat}
where
\begin{gather*}
\psi _{\pm }^{\left( \pm 1/2\right) }=\pm \big[ \psi ^{\left( \mp
1/2\right) },E_{\pm }^{(\pm 1)}\big] \in \mathcal{M}_{F}^{(\pm 1/2)}.
\end{gather*}

The RHS of (\ref{RH}) can be written in an equivalent way because we have
\[
\Theta ^{-1}(t)\Pi (t)=\Theta ^{-1}(t)BM=\big( B^{-1}\Theta \big)
^{-1}M
\]
 and this motivates the following

\begin{definition}
The gauge-equivalent relativistic sector is def\/ined by the following
set of
evolution equations%
\begin{alignat}{3}
& \partial _{+}\Theta ^{\prime }  =  +\big( \Theta ^{\prime
}E_{+}^{(+1)}\Theta ^{\prime -1}\big) _{\leq 0}\Theta ^{\prime
},\qquad && \partial _{+}\Pi ^{\prime }=-\big( \Theta ^{\prime
}E_{+}^{(+1)}\Theta ^{\prime -1}\big) _{>0}\Pi ^{\prime },&
\notag \\
& \partial _{-}\Theta ^{\prime }  = +\big( \Pi ^{\prime }E_{-}^{(-1)}\Pi
^{\prime -1}\big) _{\leq 0}\Theta ^{\prime },\qquad && \partial _{-}\Pi ^{\prime }=-\big( \Pi ^{\prime }E_{-}^{(-1)}\Pi
^{\prime -1}\big) _{>0}\Pi ^{\prime },  & \label{Gauge isospectral evolution}
\end{alignat}%
where $\Theta ^{\prime }=B^{-1}\Theta $ and $\Pi ^{\prime }=M.$ The
$\left( \ast \right) _{>0}$ denote projection onto grades $\geq
+1/2$ and $\left( \ast \right) _{\leq 0}$ onto grades $\leq 0$.
\end{definition}

The Lax covariant derivative extracted from (\ref{Gauge isospectral
evolution}) has a Lax connection
\begin{alignat*}{3}
& L_{-}^{\prime }  = \partial _{-}+A_{-}^{\prime L},\qquad &&
A_{-}^{\prime L}=B^{-1}\partial _{-}B-\psi _{-}^{\left( -1/2\right)
}-E_{-}^{(-1)},   & \notag\\
& L_{+}^{\prime }  = \partial _{+}+A_{+}^{\prime L},\qquad &&
A_{+}^{\prime L}=B^{-1}\big( E_{+}^{(+1)}+\psi _{+}^{\left(
+1/2\right) }\big) B & %\label{gauge Lax pair}
\end{alignat*}%
and it is related to (\ref{Lax pair}) by a gauge transformation $%
L\rightarrow L^{\prime }$, where $A^{\prime
L}=B^{-1}A^{L}B-dB^{-1}B$. Clearly, the two def\/initions are
equivalent.

The constant part of the Lax connection is given by ($\Sigma $ is
parametrized by $x^{\pm }$)
\begin{gather*}
E^{(\pm 1)}=E_{\pm }^{(\pm 1)}dx^{\pm }\in \Omega _{B}\left( \Sigma
\right) \otimes \widehat{\mathfrak{g}}_{(\pm 1)}
\end{gather*}%
and change under coordinate transformations because of their
$dx^{\pm }$ basis. We also have that
\begin{equation*}
\psi _{\pm }^{\left( \pm 1/2\right) }dx^{\pm }=\Omega _{F}\left(
\Sigma \right) \otimes \widehat{\mathfrak{g}}_{(\pm 1/2)}
\end{equation*}%
are fermionic 1-forms. Thus, $A^{L}$ is a superalgebra-valued
1-form. This is to recall that no superspace formulation is involved
in the construction of our super-Lax operators and that the approach
relies entirely on pure Lie algebraic properties.

The equations of motion are def\/ined by the zero curvature of
$A^{L},$ namely $\left[ L_{+},L_{-}\right] =0$ and leads to a system
of non-linear dif\/ferential equations in which the derivatives
$\partial _{\pm }$ appear mixed with the same order, hence the
name relativistic. The coupling of one positive and one negative
higher graded isospectral f\/low of opposite sign is direct from the
construction. This allows the construction of relativistic-like
integrable equations, see equation~(\ref{higher sinh-gordon equation})
below for an example.

In the def\/initions of the Lax operators above we actually have
\begin{gather}
-\partial _{+}BB^{-1}  = A_{+}^{(0)}+Q_{+}^{(0)},   \qquad
-B^{-1}\partial _{-}B  = A_{-}^{(0)}+Q_{-}^{(0)},  \label{current relation}%\notag
\end{gather}
where (the upper label denoting the $Q$ grading)
\begin{gather*}
A_{\pm }^{(0)}  = \pm \big[ \psi ^{(\mp 1)},E_{\pm }^{(\pm
1)}\big] \in
\mathcal{M}_{B}^{(0)}, \qquad
Q_{\pm }^{(0)}  = \frac{1}{2}\big[ \psi ^{(\mp 1/2)},\big[ \psi
^{(\mp 1/2)},E_{\pm }^{(\pm 1)}\big] \big] \in
\mathcal{K}_{B}^{(0)}.
\end{gather*}%
These relations are the solutions to the grade~$-1$ and~$+1$
components of the zero curvature relations $\left[
L_{+},L_{-}\right] _{-1}=\left[ L_{+}^{\prime },L_{-}^{\prime
}\right] _{+1}=0$ for the operators $L_{\pm }$
and $L_{\pm }^{\prime }$ obtained from~(\ref{isospectral evolution}) and~(\ref{Gauge isospectral evolution}).

The presence of the fermion bilinear $Q_{\pm }^{(0)}$ results in the
non-locality of the odd $t_{\pm 1/2}$ symmetry f\/lows~\cite{half-integer gradations} and also in the existence of gauge
symmetries of the models as can be deduced from the of\/f-shell
formulation of the system~(\ref{Lax pair}) done in~\cite{yo}. Having
$\mathcal{K}_{B}^{(0)}\neq \varnothing $ translates into the
existence of f\/lat directions of the Toda potential which takes the
models out of the mKdV hierarchy. Thus, we impose the vanishing
of $Q_{\pm }^{(0)}.$ Another reason why we
impose $Q_{\pm
}^{(0)}=0,$ is to get a well def\/ined relation between the dressing
matrix $\Theta $ and the term $A_{\pm }^{(0)}$ in the
spirit of~\cite{symmetry flows}, which means that the dynamical
f\/ields are described entirely in terms of the image part of the
algebra $\mathcal{M}$. The kernel part $\mathcal{K}$ is responsible
only for the symmetries of the model and all this together clarif\/ies
the role played by the term $Q_{\pm }^{(0)}.$ Then, by restricting
to superalgebras in which $Q_{\pm }^{(0)}=0$ we have local $t_{\pm
1/2}$ f\/lows and models inside the mKdV hierarchy.

\begin{remark}
Flat directions in the Toda potential $V_{B}=\big\langle
E_{+}^{(+1)}BE_{-}^{(-1)}B^{-1}\big\rangle $ allows the existence
of soliton solutions with Noether charges, e.g.\ the electrically
charged solitons of the complex sine-Gordon model which is known to
belong to the relativistic sector of the AKNS hierarchy~\cite{AKNS}
instead of the mKdV.
\end{remark}

We parametrize the Toda f\/ield as $B=g\exp [\eta Q]\exp [\nu C],$
provided we have a subalgebra solution to the algebraic conditions
$Q_{\pm }^{(0)}=0$. The model is then def\/ined on a reduced group
manifold and (\ref{current
relation}) is conveniently parametrized in the image part $\mathcal{M}%
_{B}^{(0)} $ of the algebra, i.e.\ $-\partial _{+}BB^{-1}=A_{+}^{(0)}$ and $%
-B^{-1}\partial _{-}B=A_{-}^{(0)}$.

The zero curvature $\left( F_{L}=0\right) $ of (\ref{Lax pair})
gives the supersymmetric version of the Leznov--Saveliev equations~\cite{half-integer
gradations}%
\begin{gather}
0  = F_{+-}^{(+1/2)}=-\partial _{-}\psi _{+}^{\left( +1/2\right)
}+\big[
B\psi _{-}^{\left( -1/2\right) }B^{-1},E_{+}^{(+1)}\big],  \notag \\
0  = F_{+-}^{(0)}=\partial _{-}\big( \partial _{+}BB^{-1}\big)
-\big[ E_{+}^{(+1)},BE_{-}^{(-1)}B^{-1}\big] -\big[ \psi
_{+}^{\left(
+1/2\right) },B\psi _{-}^{\left( -1/2\right) }B^{-1}\big],  \notag \\
0  = F_{+-}^{(-1/2)}=B\big( -\partial _{+}\psi _{-}^{\left(
-1/2\right) }+ \big[ E_{-}^{(-1)},B^{-1}\psi _{+}^{\left(
+1/2\right) }B\big] \big) B^{-1}.  \label{SUSY-Toda
equations}
\end{gather}

Written more explicitly in the form%
\begin{gather*}
\partial _{-}\psi _{+}^{\left( +1/2\right) }  = e^{-\eta /2}\big[ g\psi
_{-}^{\left( -1/2\right) }g^{-1},E_{+}^{(+1)}\big],  \\
\partial _{-}\left( \partial _{+}gg^{-1}\right) +\partial _{-}\partial
_{+}\nu C  = e^{-\eta }\big[
E_{+}^{(+1)},gE_{-}^{(-1)}g^{-1}\big] +e^{-\eta /2}\big[
\psi _{+}^{\left( +1/2\right) },g\psi _{-}^{\left(
-1/2\right) }g^{-1}\big],  \\
\partial _{+}\psi _{-}^{\left( -1/2\right) } =e^{-\eta /2}\big[
E_{-}^{(-1)},g^{-1}\psi _{+}^{\left( +1/2\right) }g\big],  \qquad
\partial _{-}\partial _{+}\eta Q =0,
\end{gather*}%
the linearized equations of motion with $\eta =\eta _{0}$, $\eta
_{0}\in
\mathbb{R}
$ may be written in the Klein--Gordon form
\begin{gather*}
\left( \partial _{+}\partial _{-}+\widehat{m}^{2}\right) \circ (\Xi ) = 0,\qquad
\partial _{+}\partial _{-}\nu -\Lambda  =0,
\end{gather*}
for $\Xi =\psi ^{\left( \pm 1/2\right) }$ and $\log g$, where $\widehat{m}%
^{2}$ is the mass operator%
\begin{gather*}
\widehat{m}^{2}(\Xi )=e^{-\eta _{0}}\big( {\rm ad}\,E_{-}^{(-1)}\circ {\rm ad}\, E_{+}^{(+1)}\big) \circ (\Xi )=m^{2}I\Xi.
\end{gather*}%
We have used $e^{-\eta _{0}}\big[ E_{+}^{(+1)},E_{-}^{\left(
-1\right) }\big] =\Lambda C.$ Then, the Higgs-like f\/ield $\eta
_{0}$ sets the mass scale of the theory. The massless limit
corresponds to $\eta _{0}\rightarrow \infty $. Note that all f\/ields
have the same mass which is what we would expect in a supersymmetric
theory. Taking $\eta =\eta _{0}$,
the free fermion equations of motion reads
\begin{gather*}
\partial _{\pm }\psi ^{\left( \pm 1/2\right) }=\mp \widehat{m}^{\pm }(\psi
^{\left( \mp 1/2\right) }),
\end{gather*}%
where $\widehat{m}^{\pm }(\ast )=e^{-\eta _{0}/2}{\rm ad}\,E_{\pm
}^{(\pm 1)}\circ (\ast ).$ These equations show that
fermions of opposite `chirality' are mixed by the mass term and that
in the massless limit they decouple. This means that
positive/negative f\/lows are naturally related to the two chiralities
in the f\/ield theory. In most of the literature, only the positive
set of times is usually considered.

The role of the f\/ields $\nu $ and $\eta $ associated to the central
term $C$ (of the Kac--Moody algebra~$\widehat{\mathfrak{g}}$) and
grading operator $Q$ is to restore the conformal symmetry of the
models associated to the Loop algebra $\widetilde{\mathfrak{g}}$
(which are non-conformal) so we are actually dealing with conformal
af\/f\/ine Toda models.

\subsection[Non-Abelian flows: the odd Lax pairs $L_{\pm 1/2}$]{Non-Abelian f\/lows: the odd Lax pairs $\boldsymbol{L_{\pm 1/2}}$}\label{section2.3}

Here we deduce the two lowest odd degree fermionic Lax operators
giving rise to the $\pm 1/2$ supersymmetry f\/lows, which are the ones
we are mainly concerned in the body of the paper. The negative part
is the novelty here. From (\ref{positive variarion}) and
(\ref{negative variation}) we have

\begin{definition}
The non-Abelian evolution equations of the Dressing matrices are
def\/ined by
\begin{alignat}{3}
& \delta _{K^{(+)}}\Theta  = -\big( \Theta K^{(+)}\Theta ^{-1}\big)
_{<0}\Theta,\qquad && \delta _{K^{(+)}}\Pi =+\big( \Theta
K^{(+)}\Theta ^{-1}\big) _{\geq 0}\Pi,  & \notag\\
& \delta _{K^{(-)}}\Theta  = -\big( \Pi K^{(-)}\Pi ^{-1}\big)
_{<0}\Theta,\qquad && \delta _{K^{(-)}}\Pi =+\big( \Pi
K^{(-)}\Pi ^{-1}\big) _{\geq 0}\Pi, &  \label{non-abelian flows}
\end{alignat}
for some positive/negative degree generators $K^{(+)}$ and $K^{(-)}$
in the
kernel of the operators ${\rm Ad}\, E^{(\pm 1)}$. Equivalently, we have%
\begin{alignat}{3}
& \delta _{K^{(+)}}\Theta ^{\prime }  = -\big( \Theta ^{\prime
}K^{(+)}\Theta ^{\prime -1}\big) _{\leq 0}\Theta ^{\prime },\qquad &&
\delta _{K^{(+)}}\Pi ^{\prime }=+\big( \Theta ^{\prime
}K^{(+)}\Theta ^{\prime
-1}\big) _{>0}\Pi ^{\prime }, & \notag  \\
& \delta _{K^{(-)}}\Theta ^{\prime }  = -\big( \Pi ^{\prime
}K^{(-)}\Pi ^{\prime -1}\big) _{\leq 0}\Theta ^{\prime },\qquad && \delta _{K^{(-)}}\Pi ^{\prime }=+\big( \Pi ^{\prime
}K^{(-)}\Pi ^{\prime -1}\big) _{>0}\Pi ^{\prime }. & \label{gauge non-abelian flows}
\end{alignat}
\end{definition}

The consistency of all f\/lows, as an algebra, is encoded in the
following

\begin{proposition}\label{proposition1}
The flows \eqref{non-abelian flows} and \eqref{gauge non-abelian
flows}
satisfy%
\begin{gather*}
\big[ \delta _{K_{i}^{(\pm )}},\delta _{K_{j}^{(\pm )}}\big]
(\ast )
 = \delta _{\mp \big[ K_{i}^{(\pm )},K_{j}^{(\pm )}\big] }(\ast ),\qquad
\big[ \delta _{K_{i}^{(+)}},\delta _{K_{j}^{(-)}}\big] (\ast )
 = 0,
\end{gather*}
where $(\ast )=\Theta $, $\Pi $, $\Theta ^{\prime }$, $\Pi ^{\prime
}$. The map $\delta :\mathcal{K\rightarrow \delta }_{\mathcal{K}}$
is a homomorphism.
\end{proposition}

\begin{proof}
The proof is straightforward after noting that $\left[ X,Y\right] _{\pm }=
\left[ X_{\pm },Y_{\pm }\right] +\left[ X_{\pm },Y_{\mp }\right] _{\pm }+ \left[ X_{\mp },Y_{\pm }\right] _{\pm }$.
\end{proof}

The last relation above means that the symmetries generated by elements in $\mathcal{K}_{\pm }$ commute themselves. This can be traced back to
be a consequence of the second Lie structure induced on
$\widehat{\mathfrak{g}}$
by the action of the dressing group which introduces a classical super $r$-matrix $R=\frac{1}{2}\left( \mathcal{P}_{+}-\mathcal{P}_{-}\right)
$ def\/ined in terms of the projections $\mathcal{P}_{+}$ and
$\mathcal{P}_{-}$
of $\widehat{\mathfrak{g}}=\widehat{\mathfrak{g}}_{+}+\widehat{\mathfrak{g}}%
_{-}$ along the positive/negative subalgebras
$\widehat{\mathfrak{g}}_{\pm }$,  see also \cite{Delduc-Gallot}. The
map $\delta :\mathcal{K\rightarrow \delta }_{\mathcal{K}}$ is
actually a map (up to a global irrelevant sign)
to the $R$-bracket (see \cite{BBT}) $\left[ \delta _{K},\delta _{K'}
\right] =\delta _{\left[ K,K'\right]_{R}},$ where $\left[ K,K'\right] _{R}=\left[ K,R(K')\right] +\left[ R(K),K'\right] $. Hence, all the symmetries generated by $\mathcal{K}$ are
chiral as a consequence of the second Lie structure. In particular,
this imply the commutativity of the 2D rigid supersymmetry
transformations cf.~(\ref{homo}) below, as expected.

The $\pm 1/2$ f\/lows are generated by the elements $\mp D^{(\pm
1/2)}\in \mathcal{K}_{F}^{(\pm 1/2)}$ of grades $\pm 1/2$ in the
fermionic part of the kernel, where $D^{(\pm 1/2)}$ depend on the
inf\/initesimal constant grassmannian parameters. They def\/ine the
evolution equations (actually variations cf.~(\ref{positive variarion}), (\ref{negative variation}))
\begin{alignat*}{3}
& \delta _{+1/2}\Theta  = +\big( \Theta D^{(+1/2)}\Theta ^{-1}\big)
_{<0}\Theta, \qquad &&  \delta _{+1/2}\Pi =-\big( \Theta
D^{(+1/2)}\Theta ^{-1}\big) _{\geq 0}\Pi, & \notag
 \\
& \delta _{-1/2}\Theta  = -\big( \Pi D^{(-1/2)}\Pi ^{-1}\big)
_{<0}\Theta,\qquad && \delta _{-1/2}\Pi =+\big( \Pi
D^{(-1/2)}\Pi ^{-1}\big) _{\geq 0}\Pi,&  %\label{susy variation on dressing matrices}
\end{alignat*}
giving rise to the dressing expressions
\begin{gather*}
\Theta \big( \delta _{+1/2}+D^{(+1/2)}\big) \Theta ^{-1}
 = \delta
_{+1/2}+D^{(0)}+D^{(+1/2)}=L_{+1/2}, \\
\Pi \big( \delta _{-1/2}+D^{(-1/2)}\big) \Pi ^{-1}  = \delta
_{-1/2}+BD^{(-1/2)}B^{-1}=L_{-1/2},
\end{gather*}%
where $D^{(0)}=\big[ \psi ^{\left( -1/2\right) },D^{(+1/2)}\big]
\in \mathcal{M}_{B}^{(0)}.$ The derivation of $L_{-1/2}$ follows
exactly the same lines for the derivation of $L_{+1/2}$ done in
\cite{half-integer
gradations}. At this point we have four Lax operators~$L_{\pm 1/2}$ and~$L_{\pm 1}$. The grade subspace decomposition of the relations
$\left[ L_{\pm 1/2},L_{+1}\right] =\left[ L_{\pm 1/2},L_{-1}\right]
=0$ allows to take the solution $D^{(0)}=-\delta _{+1/2}BB^{-1}.$
The compatibility of this system of four Lax operators provides the
2D supersymmetry transformations among the f\/ield components. Indeed,
using the equations of motion we get their explicit form, see equation~(\ref{susy transformations 1/2}) and (\ref{susy transformations
-1/2}) below.

Finally, the odd Lax operators reads
\begin{gather}
L_{+1/2} =\delta _{+1/2}-\delta _{+1/2}BB^{-1}+D^{(+1/2)},
\label{Lax +1/2}
\\
L_{-1/2} =\delta _{-1/2}+BD^{(-1/2)}B^{-1}.  \label{Lax -1/2}
\end{gather}%
The operator $L_{+1/2}$ was already constructed in
\cite{half-integer gradations} and the $L_{-1/2}$ is the novelty
here.

Note that in (\ref{Lax +1/2}) and (\ref{Lax -1/2}) are in dif\/ferent
gauges. This is the key idea for introducing the Toda potential
(superpotential) in the supersymmetry transformations which is also
responsible for coupling the two sectors.

\subsection[Local supersymmetry flows $\delta _{\pm 1/2}$]{Local supersymmetry f\/lows $\boldsymbol{\delta _{\pm 1/2}}$}\label{section2.4}

The equations (\ref{SUSY-Toda equations}) are invariant under a pair
of non-Abelian fermionic f\/lows $(\delta _{\rm SUSY}=\delta
_{-1/2}+\delta _{+1/2})$
as a consequence of the compatibility relations $\left[ L_{\pm 1/2},L_{+}%
\right] =\left[ L_{\pm 1/2},L_{-}\right] =0$ supplemented by the
equations of motion $\left[ L_{+},L_{-}\right] =0$ and the Jacobi
identity. They are generated by the elements in the fermionic kernel
$\mathcal{K}_{F}^{(\pm 1/2)}$ and are explicitly given by
\begin{gather}
\delta _{+1/2}\psi _{-}^{\left( -1/2\right) } =\big[
E_{-}^{(-1)},B^{-1}D^{(+1/2)}B\big],  \qquad
\delta _{+1/2}BB^{-1} =\big[ D^{(+1/2)},\psi ^{\left( -1/2\right)
}\big],
\notag \\
\delta _{+1/2}\psi _{+}^{\left( +1/2\right) } =\big[ \delta
_{+1/2}BB^{-1},\psi _{+}^{\left( +1/2\right) }\big] -\big[
\partial _{+}BB^{-1},D^{(+1/2)}\big].   \label{susy transformations 1/2}
\end{gather}
and
\begin{gather}
\delta _{-1/2}\psi _{-}^{\left( -1/2\right) }  = -\big[
B^{-1}\delta _{-1/2}B,\psi _{-}^{\left( -1/2\right) }\big] -\big[
B^{-1}\partial
_{-}B,D^{(-1/2)}\big],   \notag \\
B^{-1}\delta _{-1/2}B  = \big[ D^{(-1/2)},\psi ^{\left( +1/2\right)
}\big],
\qquad
\delta _{-1/2}\psi _{+}^{\left( +1/2\right) }  = \big[ E_{+}^{(+1)},BD^{(-1/2)}B^{-1}\big]. \label{susy transformations -1/2}
\end{gather}%
The physical degrees of freedom are parametrized by the image part $\mathcal{M}$. To guarantee that the variations of the f\/ields remain in~$\mathcal{M}$ we have to check that the kernel components of the above
transformations vanishes, i.e.
\begin{equation}
\big[ \delta _{+1/2}BB^{-1},\psi _{+}^{\left( +1/2\right) }\big]
\in \mathcal{K}=0,\qquad \big[ B^{-1}\delta _{-1/2}B,\psi
_{-}^{\left( -1/2\right) }\big] \in \mathcal{K}=0.
\label{vanishing condition}
\end{equation}%
We will see below in the examples that $Q_{\pm }^{(0)}$ $=0$ imply (\ref{vanishing condition}) as a consequence of the absence of the even graded $(2n,n\in \mathbb{Z}
)$ part of the bosonic kernel $\mathcal{K}_{B}$ in the mKdV
hierarchy. These conditions turn the lowest odd f\/lows $\delta
_{\pm 1/2}$ local.

The Lax operators (\ref{Lax +1/2}), (\ref{Lax -1/2}) generating the
odd f\/lows (\ref{susy transformations 1/2}), (\ref{susy transformations -1/2}) are related to the rigid 2D supersymmetry
transformations of the type
\begin{equation*}
N=\left( N_{+},N_{-}\right),  %\label{N}
\end{equation*}
where $N_{\pm }=\dim \mathcal{K}_{F}^{(\pm 1/2)}$. As the map
$D^{(\pm 1/2)}\rightarrow \delta _{\pm 1/2}$ obeys
\begin{gather}
\big[ \delta _{\pm 1/2},\delta'_{\pm 1/2}\big] (\ast )  = \partial _{\mp \left[ D^{\pm 1/2},D'{}^{\pm 1/2}\right] }(\ast )\sim \partial _{\pm }(\ast ),   \qquad
\big[ \delta _{+1/2},\delta _{-1/2}\big] (\ast ) = 0,  \label{homo}
\end{gather}
we see that two fermionic transformations close into derivatives,
which is by def\/inition a supersymmetry. This is the case provided
$\frac{1}{2}\left\{ F^{(\pm 1/2)},F^{(\pm 1/2)}\right\} \sim E_{\pm
}^{(\pm 1)}$ for $F^{(\pm 1/2)}\in \mathcal{K}_{F}^{(\pm 1/2)},$
which is signif\/icant for the supersymmetric structure of the models,
see for instance~\cite{Inami-Kanno}. For simplicity, we take
constant elements $E_{\pm }^{(\pm 1)}$ which are dual $\big(
E_{+}^{(+1)}\big) ^{\ast }=E_{-}^{(-1)},$ giving rise to
isomorphic subspaces $\mathcal{K}_{F}^{(+1/2)}\simeq
\mathcal{K}_{F}^{(-1/2)}
$ and to $N_{+}=N_{-}$ in consistency with the pairing induced by $%
\left\langle K_{i},K_{j}\right\rangle \sim \delta _{i+j}.$ Note that
the non-Abelian odd f\/lows close into the isospectral even f\/lows, as
expected, and that the central and gradation f\/ields do not transform
under $\delta _{\pm 1/2}$ then, they are not truly degrees of
freedom of the model.

\subsection{Recursion operators and higher odd f\/lows}\label{section2.5}

In computing the explicit expressions for odd Lax operators using (\ref{non-abelian flows}) generating higher degree fermionic f\/lows we realize that this is considerably more involved than the $\pm 1/2$ cases. Instead of
that, we use the dressing map $\mathcal{K\rightarrow \delta
}_{\mathcal{K}}$ from the kernel algebra to the f\/low algebra in
order to introduce recursion operators. From the relations
\begin{gather*}
\left[ \mathcal{\delta }_{K^{(\pm 1)}},\mathcal{\delta }_{F^{(\pm 1/2)}}
\right] (\ast )=\mathcal{\delta }_{\left[ K^{(\pm 1)},F^{(\pm
1/2)}\right] }(\ast )=\delta _{F^{(\pm 3/2)}}(\ast ),
\end{gather*}
we infer the following behavior
\begin{gather*}
\mathcal{\delta }_{F^{(\pm n\pm 1/2)}}(\ast )  = {\rm ad}_{\mathcal{\delta }
_{K^{(\pm 1)}}}^{n}\left( \mathcal{\delta }_{F^{(\pm 1/2)}}\right)
(\ast )=\left( \mathcal{R}^{\pm 1}\right) ^{n}\left( \mathcal{\delta
}_{F^{(\pm
1/2)}}\right) (\ast ),   \notag \\
\mathcal{R}^{\pm 1}\mathcal{(\ast )}  ={\rm ad}_{\mathcal{\delta }
_{K^{(\pm 1)}}}(\ast )=\left[ \mathcal{\delta }_{K^{(\pm 1)}},\mathcal{\ast }
\right] %\label{recursion}
\end{gather*}
in terms of the recursion operators $\mathcal{R}^{\pm 1}$. The aim
is not to reproduce the well known supersymmetry transformations but
to develop a method to construct systematically all the Higher graded
odd symmetry f\/lows in terms of its simplest symmetry structure.
However, we have to recognize that the use of super
pseudo-dif\/ferential operators and associated scalar Lax operators,
seems to be more appropriated for computational purposes.

From this analysis, we have the following two chains of
supersymmetry
transformations
\begin{gather}
 \mathcal{\delta }_{+1/2}\overset{\delta _{K^{+}}}{\rightarrow }\mathcal{\delta }_{+3/2}\overset{\delta _{K^{+}}}{\rightarrow }\mathcal{\delta }
_{+5/2}\overset{\delta _{K^{+}}}{\rightarrow }\mathcal{\delta
}_{+7/2} \overset{\delta _{K^{+}}}{\rightarrow } \cdots,
\notag \\
\cdots \overset{\delta _{K^{-}}}{\leftarrow }  \mathcal{\delta }_{-7/2}\overset{\delta _{K^{-}}}{\leftarrow }\mathcal{\delta }_{-5/2}\overset{\delta _{K^{-}}}{\leftarrow }\mathcal{\delta }_{-3/2}
\overset{\delta _{K^{-}}}{\leftarrow }\mathcal{\delta }_{-1/2},\label{flows}
\end{gather}
where the ones corresponding to $\mathcal{\delta }_{\pm 1/2}$ are
considered
as starting points. The variations $\delta _{K^{\pm }}$ are given by (\ref{non-abelian flows}) or (\ref{gauge non-abelian flows}). For
example, for a degree $+1$ element $K^{(+1)}$ we have from
(\ref{non-abelian flows}) that
\begin{gather}
\delta _{K^{(+1)}}\psi _{+}^{\left( +1/2\right) }  = -\big[
E_{+}^{(+1)},\big( \Theta K^{(+1)}\Theta ^{-1}\big) _{-1/2}\big|_{\mathcal{M}}
\big], \notag \\
\delta _{K^{(+1)}}\big( \partial _{+}BB^{-1}\big)  = +\big[
E_{+}^{(+1)},\big( \Theta K^{(+1)}\Theta ^{-1}\big) _{-1}\big]
+\big[ \psi _{+}^{\left( +1/2\right) },\big( \Theta K^{(+1)}\Theta
^{-1}\big)
_{-1/2}\big|_{\mathcal{K}}\big],  \notag \\
\delta _{K^{(+1)}}\psi _{-}^{\left( -1/2\right) }  = -\big[
E_{-}^{(-1)},B^{-1}\big( \Theta K^{(+1)}\Theta ^{-1}\big) _{+1/2}B\big|_{\mathcal{M}}\big].
\label{K+ transformations}
\end{gather}

The dressing matrix $\Theta $ factorizes as $\Theta =US,$ where
$U\in \exp \mathcal{M}$ is local and $S\in \exp \mathcal{K}$ is
non-local in the f\/ields \cite{half-integer gradations}, splitting
the Dressing of the vacuum Lax operators $\left( L_{\pm }=\Theta
L_{\pm }^{V}\Theta ^{-1}\right) $ as a two step process. A $U$ and
an $S$ rotation given respectively by
\begin{gather}
U^{-1}L_{+}U  = \partial _{+}+E_{+}^{(+1)}+K_{+}^{(-)},\qquad
U^{-1}L_{-}U=\partial _{-}+K_{-}^{(-)},  \label{u} \\
S^{-1}\big( \partial _{+}+E_{+}^{(+1)}+K_{+}^{(-)}\big) S
 = \partial _{+}+E_{+}^{(+1)},\qquad S^{-1}\big( \partial
_{-}+K_{-}^{(-)}\big) S=\partial _{-}+E_{-}^{(-1)},  \label{S}
\end{gather}
where $K_{\pm }^{(-)}$ $\in \mathcal{K}$ involve expansions on the
negative grades only. The components $\psi ^{(i)}$,
$i=-1/2,-3/2,\dots$ of $U$ are
extracted by projecting (\ref{u}) along $\mathcal{M}$ and the components $%
s^{(i)},$ $i=-1/2,-3/2,\dots$ by projecting (\ref{S}) along
$\mathcal{K}$. This allows to compute~(\ref{K+ transformations}).
The higher graded supersymmetry transformations are inevitably
non-local because of the presence of the kernel part $S$ appearing
in the def\/inition of the transformations $\delta _{K^{(\pm 1)}}$
used to construct them. Thus, the
best we can do is to restrict ourselves to a reduced manifold (def\/ined by $Q_{\pm }^{(0)}=0$) in which $\mathcal{\delta }_{\pm 1/2}$ are local. From (\ref{non-abelian flows}) we have
\begin{equation*}
\big[ \delta _{K^{(+1)}},\mathcal{\delta }_{-1/2}\big] (\ast )=\delta _{\left[ K^{(+1)},D^{(-1/2)}\right] _{R}}(\ast )=0
\end{equation*}
and we cannot connect $\mathcal{\delta }_{-1/2}$ and $\mathcal{\delta
}_{+1/2}$ through a $\delta _{K^{(+1)}}$ f\/low, ref\/lecting the
chiral independence of
the $\mathcal{\delta }_{\pm 1/2}$ transformations as a consequence of the $R$-bracket. This is why in~(\ref{flows}) the sectors are treated
separately. Although the higher graded odd f\/lows are non-local,
their square always give
a local even f\/low. A similar conclusion for this behavior was found in~\cite{Mathieu} by using superspace formalism.

\subsection{Generalized relativistic-like current identities}\label{section2.6}

In this section we derive an inf\/inite set of identities associated
to the f\/lows generated by $\mathcal{K}_{F}^{(\pm 1/2)}.$ The word
relativistic is used in the sense that each~$t_{\pm n}$ is coupled
to its opposite counterpart~$t_{\mp n}$.

\begin{proposition}
The infinite set of fermionic local currents defined by
\begin{alignat}{3}
& J_{+n}^{(+1/2)}  = \big\langle D^{(+1/2)}\Theta E^{(+n)}\Theta
^{-1}\big\rangle,\qquad && J_{-n}^{(+1/2)}=\big\langle
D^{(+1/2)}\Pi
E^{(-n)}\Pi ^{-1}\big\rangle,  & \notag \\
& J_{+n}^{(-1/2)}  = \big\langle D^{(-1/2)}\Theta ^{\prime
}E^{(+n)}\Theta ^{\prime -1}\big\rangle,\qquad && J_{-n}^{(-1/2)}=\big\langle D^{(-1/2)}\Pi ^{\prime }E^{(-n)}\Pi
^{\prime -1}\big\rangle, &  \label{current components}
\end{alignat}
satisfy the following identities
\begin{equation}
\partial _{+n}J_{-n}^{(\pm 1/2)}-\partial _{-n}J_{+n}^{(\pm 1/2)}=0.
\label{generalized conservation laws}
\end{equation}%
The $D^{(\pm 1/2)}\in \mathcal{K}_{F}^{(\pm 1/2)}$ are the
generators of the fermionic kernel.
\end{proposition}

\begin{proof}
The proof is extremely simple and is based only on the relations (\ref{Isospectral evolutions}) and (\ref{Isospectral II}). Start with
\begin{gather*}
\partial _{+n}J_{-m}^{(+1/2)}  = \big\langle D^{(+1/2)}\big[ \big( \Theta
E^{(+n)}\Theta ^{-1}\big) _{\geq 0},\big( \Pi E^{(-m)}\Pi
^{-1}\big)
_{<0}\big] \big\rangle, \\
\partial _{-n}J_{+m}^{(+1/2)}  = \big\langle D^{(+1/2)}\big[ \big( \Theta
E^{(+m)}\Theta ^{-1}\big) _{\geq 0},\big( \Pi E^{(-n)}\Pi
^{-1}\big) _{<0}\big] \big\rangle
\end{gather*}%
to get%
\begin{gather*}
\partial _{+n}J_{-n}^{(\pm 1/2)}-\partial _{-n}J_{+n}^{(\pm 1/2)}
 = \big\langle D^{(+1/2)}\big[ \big( \Theta E^{(+n)}\Theta
^{-1}\big)
_{\geq 0},\big( \Pi E^{(-m)}\Pi ^{-1}\big) _{<0}\big] \big\rangle   \\
 \phantom{\partial _{+n}J_{-n}^{(\pm 1/2)}-\partial _{-n}J_{+n}^{(\pm 1/2)}=}{}
 -\big\langle D^{(+1/2)}\big[ \big( \Theta E^{(+m)}\Theta
^{-1}\big) _{\geq 0},\big( \Pi E^{(-n)}\Pi ^{-1}\big)
_{<0}\big] \big\rangle.
\end{gather*}%
This sum vanishes for $m=n$. For $J^{(-1/2)}$ the proof is
analogous.
\end{proof}

These identities mixes the two sectors corresponding to positive and
negative isospectral times in a relativistic manner. They can be
written in a covariant form $\eta ^{ij}\frac{\partial }{\partial
t_{i}}J_{j}^{(\pm 1/2)}=0$ if we def\/ine a constant `metric' $\eta
=\eta _{ij}dt_{i}dt_{j}$ for each pair of positive/negative times.
However, the interpretation of this higher graded `light-cone
coordinates' deserves further study.

Consider now the lowest isospectral f\/lows $t_{\pm 1}=-x^{\pm }.$ The
current components (\ref{current components}) are given by
\begin{alignat*}{3}
& J_{+}^{(+1/2)}  = -\big\langle D^{(+1/2)}\big[ \psi
^{(-1/2)},\partial _{+}BB^{-1}\big] \big\rangle,\qquad&&
J_{-}^{(+1/2)}=+\big\langle
D^{(+1/2)}B\psi _{-}^{(-1/2)}B^{-1}\big\rangle,&   \notag  \\
& J_{+}^{(-1/2)}  = +\big\langle D^{(-1/2)}B^{-1}\psi
_{+}^{(+1/2)}B\big\rangle,\qquad &&
  J_{-}^{(-1/2)}=+\big\langle D^{(-1/2)}\big[ \psi
^{(+1/2)},B^{-1}\partial _{-}B\big] \big\rangle.&  %\label{SUSY currents}
\end{alignat*}%
Then, there are $N=\dim \mathcal{K}_{F}^{(\pm 1/2)}$ associated
relativistic conservation laws (for each sector) given by $\partial
_{+}J_{-}^{(\pm 1/2)}-\partial _{-}J_{+}^{(\pm 1/2)}=0$. More
explicitly, we have
\begin{gather}
\partial _{-}\big( \big[ \psi ^{(-1/2)},\partial _{+}BB^{-1}\big] \big|_{%
\mathcal{K}}\big) +\partial _{+}\big( B\psi _{-}^{(-1/2)}B^{-1}\big|_{%
\mathcal{K}}\big)   = 0,  \notag \\
-\partial _{-}\big( B^{-1}\psi _{+}^{(+1/2)}B\big|_{\mathcal{K}}\big)
+\partial _{+}\big( \big[ \psi ^{(+1/2)},B^{-1}\partial _{-}B\big] \big|_{%
\mathcal{K}}\big)   = 0.  \label{SUSY conservation laws}
\end{gather}%
This time, the identities provide supercharge conservation laws due
to the fact that the f\/lows $t_{\pm 1}$ are identif\/ied with the
light-cone coordinates $x^{\pm }=\frac{1}{2}\left( x^{0}\pm
x^{1}\right) $. It is not clear if the identities associated to the
higher f\/lows $t_{\pm n}$, $n\geq +1$ provide new conserved
quantities because one is not supposed to impose boundary conditions
or integrate along these directions. For higher times they are taken
as simple identities consequence of the f\/low relations above.

Now that we have $N=\dim \mathcal{K}_{F}^{(\pm 1/2)}$ supercurrents
associated to $\mathcal{K}_{F}^{(\pm 1/2)},$ let's compute their
corresponding supercharges by the Noether procedure in order to
check that they really generate the supersymmetry transformations
(\ref{susy transformations 1/2}) and (\ref{susy transformations
-1/2}).

\subsection[Supercharges for the SUSY flows $\delta _{\pm 1/2}$]{Supercharges for the SUSY f\/lows $\boldsymbol{\delta _{\pm 1/2}}$}\label{section2.7}

The action for the af\/f\/ine supersymmetric Toda models was deduced in
\cite{yo} and it is given by
\begin{gather}
S_{\rm Toda}[B,\psi ]  = S_{\rm WZNW}[B]-\frac{k}{4\pi }\int_{\Sigma
}\big\langle \psi _{+}^{\left( +1/2\right) }\partial _{-}\psi
^{\left( -1/2\right) }+\psi _{-}^{\left( -1/2\right) }\partial
_{+}\psi ^{\left( +1/2\right)
}\big\rangle   \notag \\
\phantom{S_{\rm Toda}[B,\psi ]  =}{}
+\frac{k}{2\pi }\int_{\Sigma }\big\langle
E_{+}^{(+1)}BE_{-}^{(-1)}B^{-1}+\psi _{+}^{\left( +1/2\right) }B\psi
_{-}^{\left( -1/2\right) }B^{-1}\big\rangle.  \label{SUSY-Toda
action}
\end{gather}%
This corresponds to the situation when we restrict to the
sub-superalgebras solving the condition $Q_{\pm }^{(0)}=0.$ In this
case the potential ends at the second term providing a Yukawa-type
term turning the model integrable and supersymmetric. The light-cone
notation used for the f\/lat Minkowski
space $\Sigma $ is $x^{\pm }=\frac{1}{2}\left( x^{0}\pm x^{1}\right) $, $
\partial _{\pm }=\partial _{0}\pm \partial _{1}$, $\eta _{+-}=\eta
_{-+}=2$, $\eta ^{+-}=\eta ^{-+}=\frac{1}{2}$, $\epsilon _{+-}=-\epsilon
_{-+}=2$, $\epsilon ^{-+}=-\epsilon ^{+-}=\frac{1}{2}$ corresponding to
the metric $\eta _{00}=1$, $\eta _{11}=-1$ and antisymmetric symbol
$\epsilon
_{10}=-\epsilon _{01}=+1$. A coupling constant is introduced by setting $%
E_{\pm }^{(\pm 1)}\rightarrow \mu E_{\pm }^{(\pm 1)}$ and $\psi ^{\left( \pm 1/2\right) }\rightarrow \mu ^{-1/2}\psi
^{\left( \pm 1/2\right) }$.

An arbitrary variation of the action (\ref{SUSY-Toda action}) is given by%
\begin{equation*}
\frac{2\pi }{k}\delta S_{\rm Toda}=\int_{\Sigma }\big\langle \delta
BB^{-1}F_{+-}^{(0)}\big\rangle -\int_{\Sigma }\big\langle \delta
\psi ^{\left( +1/2\right) }B^{-1}F_{+-}^{(-1/2)}B\big\rangle
-\int_{\Sigma }\big\langle \delta \psi ^{\left( -1/2\right)
}F_{+-}^{(+1/2)}\big\rangle
\end{equation*}%
and the equations of motions are exactly the super Leznov--Saveliev
equations, cf.~(\ref{SUSY-Toda equations}) above.

Taking $\delta \rightarrow \delta _{\rm SUSY}=\delta _{-1/2}+\delta
_{+1/2}$, using (\ref{susy transformations 1/2}), (\ref{susy
transformations -1/2}) and considering $D^{(\pm 1/2)}$ as functions
of the coordinates $x^{\pm }$,
we have the supersymmetric variation of the action%
\begin{gather*}
\frac{2\pi }{k}\delta _{\rm SUSY}S_{\rm Toda}  = \int_{\Sigma }\big\langle
\partial _{-}D^{(-1/2)}\big( B^{-1}\psi _{+}^{\left( +1/2\right)
}B\big) -\partial _{+}D^{(-1/2)}\big[ \psi ^{\left( +1/2\right)
},B^{-1}\partial _{-}B\big]
\big\rangle   \\
\phantom{\frac{2\pi }{k}\delta _{\rm SUSY}S_{\rm Toda}  =}{}
-\int_{\Sigma }\big\langle \partial _{-}D^{(+1/2)}\big[ \psi
^{\left( -1/2\right) },\partial _{+}BB^{-1}\big] +\partial
_{+}D^{(+1/2)}\big( B\psi _{-}^{\left( -1/2\right) }B^{-1}\big)
\big\rangle.
\end{gather*}
This allows to obtain two conservation laws
\begin{gather*}
0=\int_{\Sigma }\big\langle D^{(\mp 1/2)}\big( \partial
_{-}j_{+}^{\left( \pm 1/2\right) }+\partial _{+} j_{-}^{\left( \pm 1/2\right) }\big) \big\rangle,
\end{gather*}
which are exactly the ones derived by using the extended
Riemann--Hilbert
approach (\ref{generalized conservation laws}) for the lowest f\/lows (\ref{SUSY conservation laws}). Then, there are $\dim \mathcal{K}_{F}$
supercurrents and supercharges given by

f\/low $\delta _{+1/2}$:
\begin{gather}
 j_{+}^{\left( -1/2\right)
}=\big[ \psi ^{\left( -1/2\right) },\partial _{+}BB^{-1}\big]
\big|_{\mathcal{K}},\qquad j_{-}^{\left( -1/2\right) }=B\psi _{-}^{\left( -1/2\right) }B^{-1}|_{%
\mathcal{K}},   \notag \\
Q_{+}  = \int dx^{1}\big( \big[ \psi ^{\left( -1/2\right)
},\partial
_{+}BB^{-1}\big] +B\psi _{-}^{\left( -1/2\right) }B^{-1}\big)\big|_{\mathcal{K}},  \label{Q+}
\end{gather}

f\/low $\delta _{-1/2}$:
\begin{gather}
j_{+}^{\left( +1/2\right)
}=-B^{-1}\psi _{+}^{\left( +1/2\right) }B\big|_{\mathcal{K}}, \qquad
j_{-}^{\left( +1/2\right) }=\big[ \psi ^{\left( +1/2\right)
},B^{-1}\partial _{-}B\big] \big|_{\mathcal{K}},  \notag \\
Q_{-}  = \int dx^{1}\big( \big[ \psi ^{\left( +1/2\right)
},B^{-1}\partial _{-}B\big] -B^{-1}\psi _{+}^{\left( +1/2\right)
}B\big) \big|_{\mathcal{K}}.  \label{Q-}
\end{gather}

The variation above is the same when (\ref{vanishing condition}) are
zero or not, this is because all the f\/ields are def\/ined in
$\mathcal{M}$ and the kernel part does not af\/fect the variation at
all. These two ways of extracting the supercharges show a deep
relation between the algebraic dressing formalism and the
Hamiltonian reduction giving~(\ref{SUSY-Toda action}).

Now specialize the construction done above to the simplest toy
examples. The
supercharges are computed from the general formulas~(\ref{Q+}) and~(\ref{Q-}). We want to emphasize that the sub-superalgebras solving the condition $Q_{\pm }^{(0)}=0$ have no bosonic kernel $\mathcal{K}_{B}$ of degree
zero in consistency with the absence of positive even isospectral
f\/lows $t_{+2n}$ in the mKdV hierarchy.

\section{Examples}\label{section3}

These examples show how the superspace notion of supersymmetry can
be embedded consistently into the inf\/inite-dimensional f\/low
approach. The usual
SUSY transformations corresponds to the f\/low algebra spanned by the times $(t_{-1},t_{-1/2},t_{+1/2},t_{+1})$. We can have several pairs of
odd times $t_{\pm 1/2}$ depending on the dimension of
$\mathcal{K}_{F}^{(\pm 1/2)}$ as shown above.

\subsection[The $N=(1,1)$ sinh-Gordon model reloaded]{The $\boldsymbol{N=(1,1)}$ sinh-Gordon model reloaded}
\label{section3.1}

Take the $sl(2,1)_{[1]}^{(2)}$ superalgebra (see Appendix~\ref{appendixA} for
details).
The Lagrangian is
\begin{gather}
L  = -\frac{k}{2\pi }\left[ \partial _{+}\phi \partial _{-}\phi
+\psi _{-}\partial _{+}\psi _{-}+\psi _{+}\partial _{-}\psi
_{+}-V\right],
\notag \\
V =2\mu ^{2}\cosh [2\phi ]+4\mu \psi _{+}\psi _{-}\cosh [\phi ]\label{(1,1)lagrangian}
\end{gather}
and the equations of motion are%
\begin{gather}
\partial _{+}\partial _{-}\phi  = -2\mu ^{2}\sinh [2\phi ]-2\mu \psi
_{+}\psi _{-}\sinh [\phi ],  \notag \\
\partial _{-}\psi _{+}  = 2\mu \psi _{-}\cosh [\phi ],\qquad
\partial _{+}\psi _{-}  = -2\mu \psi _{+}\cosh [\phi ]. \label{t-1 equations}
\end{gather}

With $D^{(+1/2)}=\epsilon _{-}F_{2}^{(+1/2)}$, $D^{(-1/2)}=\epsilon
_{+}F_{1}^{(-1/2)}$ and $\psi _{\pm }\rightarrow \frac{1}{2}\psi
_{\pm }$, the supersymmetry f\/lows are
\begin{gather}
\delta _{\pm 1/2}\phi  = \pm \epsilon _{\mp }\psi _{\pm },
\qquad
\delta _{\pm 1/2}\psi _{\pm }  = \mp \epsilon _{\mp }\partial _{\pm
}\phi,
\qquad
\delta _{\pm 1/2}\psi _{\mp }   = 2\mu \epsilon _{\mp }\sinh [\phi
], \notag\label{(1,1) transformations}
\end{gather}
where we have used the parametrizations
\begin{gather*}
B  = \exp [\phi H_{1}],\qquad \psi ^{\left( +1/2\right) }=\psi
_{-}G_{1}^{(+1/2)},\qquad \psi ^{\left( -1/2\right) }=\psi
_{+}G_{2}^{(-1/2)}, \\
\psi _{-}^{\left( -1/2\right) }  = 2\psi _{-}G_{2}^{(-1/2)},\qquad
\psi _{+}^{\left( +1/2\right) }=-2\psi _{+}G_{1}^{(+1/2)}.
\end{gather*}

We can check (\ref{homo}) by applying the variations twice
giving
\begin{gather*}
\big[ \delta _{\pm 1/2},\delta'
_{\pm 1/2}\big]  = 2\epsilon _{\mp }\epsilon _{\mp }^{\prime
}\partial
_{\pm }, \qquad
\left[ \delta _{+1/2},\delta _{-1/2}\right]  = 0.
\end{gather*}

Then, we have two real supercharges $N=(1,1)$ because of $\dim\mathcal{K}%
_{F}^{(\pm 1/2)}=1$. They are given~by
\begin{gather*}
\delta _{\pm 1/2}  : \ \ Q^{\left( \mp 1/2\right) }=Q_{\pm
}^{1}F_{1,2}^{(\mp 1/2)}, \qquad
Q_{\pm }^{1}  = \int dx^{1}\left( \psi _{\pm }\partial _{\pm }\phi
\mp \psi _{\mp }h^{\prime }(\phi )\right),
\end{gather*}%
where $h(\phi )=2\mu \cosh [\phi ]$ and $h^{\prime }(\phi )$ is its
functional derivative respect $\phi $.

Now rotate the fermions by the phase $\exp \left( i\pi /4\right) $
in order
to write (\ref{(1,1)lagrangian}) in a more familiar form%
\begin{gather*}
L  = -\frac{k}{\pi }\left[ \frac{1}{2}\partial _{+}\phi \partial _{-}\phi +%
\frac{i}{2}\psi _{-}\partial _{+}\psi _{-}+\frac{i}{2}\psi
_{+}\partial
_{-}\psi _{+}-V\right], \\
V  = \frac{1}{2}\left( h^{\prime }(\phi )\right) ^{2}+ih^{\prime
\prime }(\phi )\psi _{+}\psi _{-}+\mu ^{2},
\end{gather*}%
which is known to be invariant under the $N=(1,1)$ superspace
transformations for a real bosonic superf\/ield. The area term comes
from squaring $h^{\prime }(\phi )=2\mu \sinh [\phi ]$.

\begin{note}
The Poisson brackets are def\/ined by
\begin{equation*}
\left\{ A,B\right\} _{\rm PB}=-(-1)^{\epsilon _{A}\epsilon _{B}}\left( \frac{%
\partial A}{\partial f}\frac{\partial B}{\partial \pi _{f}}-(-1)^{\epsilon
_{A}\epsilon _{B}}\frac{\partial A}{\partial \pi _{f}}\frac{\partial B}{%
\partial f}\right),
\end{equation*}%
where $\epsilon =1,0$ for bosonic-fermionic quantities and $\pi _{f}=\frac{\partial L}{\partial (\partial _{t}f)}.$ The Dirac bracket is def\/ined by
\begin{equation*}
\left\{ A,B\right\} _{\rm DB}=\left\{ A,B\right\} _{\rm PB}-\left\{ A,\phi
_{i}\right\} _{\rm PB}\left( C^{-1}\right) _{ij}\left\{ \phi
_{j},B\right\} _{\rm PB},
\end{equation*}%
where $C_{ij}$ =$\left\{ \phi _{i},\phi _{i}\right\} _{\rm PB}$ and
$\phi _{i}$ are the second class constraints.
\end{note}

With the Dirac brackets $\left\{ \phi,\partial _{t}\phi \right\} =1,$ $\left\{ \psi _{\pm },\psi _{\pm }\right\} =-i$ and $Q_{\pm
}^{1}f=\left\{ Q_{\pm }^{1},f\right\} $ we have, after replacing
$Q_{\pm }^{1}\rightarrow iQ_{\pm }^{1},$ the action of the
supercharges on the f\/ield components
\begin{gather*}
Q_{\pm }^{1}\phi  = -i\psi _{\pm },\qquad
Q_{\pm }^{1}\psi _{\pm }  = +\partial _{\pm }\phi,\qquad
Q_{\pm }^{1}\psi _{\mp }  = \mp h^{\prime }(\phi ).
\end{gather*}

Finally, the total f\/low can be written as $\delta _{\rm SUSY}=-i\epsilon
_{-}Q_{+}^{1}+i\epsilon _{+}Q_{-}^{1}$
\begin{gather*}
\delta _{\rm SUSY}\phi  = +i\epsilon _{-}\psi _{+}-i\epsilon _{+}\psi _{-},\qquad
\delta _{\rm SUSY}\psi _{\pm }  = \mp \epsilon _{\mp }\partial _{\pm
}\phi +\epsilon _{\pm }h^{\prime }(\phi ),
\end{gather*}%
which are the ordinary $N=(1,1)$ supersymmetry transformations
obtained by using a superspace approach.

Now, construct the $\delta _{+3/2}$ transformations starting from
$\delta
_{+1/2}$ by applying $\mathcal{\delta }_{K^{(+1)}}$ as shown in~(\ref{flows}). From \cite{half-integer gradations} we have%
\begin{gather*}
\delta _{K^{(+1)}}\phi  = -\frac{1}{2}\alpha \psi _{+}Q(x^{+}),
\qquad
\delta _{K^{(+1)}}\psi _{+}  = +\frac{1}{2}\alpha \left( \partial
_{+}\psi
_{+}-\partial _{+}\phi Q(x^{+})\right), \notag \\
\delta _{K^{(+1)}}\psi _{-}  = -\frac{1}{2}\alpha \left( \psi
_{+}h(\phi )-h^{\prime }(\phi )Q(x^{+})\right),  %\label{K+1 transformations}
\end{gather*}
where $\displaystyle Q(x^{+})=\int^{x^{+}}\left( \psi _{+}\partial _{+}\phi
\right) $. From the relation $\big[
K_{2}^{(+1)},F_{2}^{(+1/2)}\big] =F_{1}^{(+3/2)}$ we have that
$\delta _{+3/2}=\left[ \delta _{K^{(+1)}},\delta _{+1/2}\right] $
and the transformations are given by
\begin{gather}
\delta _{+3/2}\phi  = \epsilon _{-}\left( \frac{1}{2}\partial
_{+}\psi _{+}-\partial _{+}\phi Q(x^{+})+\frac{1}{2}\psi
_{+}H(x^{+})\right),\notag
\\
\delta _{+3/2}\psi _{+}  = \epsilon _{-}\left( \frac{1}{2}\partial
_{+}^{2}\phi +\partial _{+}\psi _{+}Q(x^{+})-\frac{1}{2}\partial
_{+}\phi
H(x^{+})\right),  \notag \\
\delta _{+3/2}\psi _{-}  = \epsilon _{-}\left( -\frac{1}{2}\partial
_{+}\phi h(\phi )-\psi _{+}Q(x^{+})h(\phi )+\frac{1}{2}h^{\prime
}(\phi )H(x^{+})\right), \label{3/2 variations}
\end{gather}
where $\displaystyle H(x^{+})=\int^{x^{+}}\left( \left( \partial _{+}\phi \right)
^{2}+\psi _{+}\partial _{+}\psi _{+}\right) $. We also f\/ind the variations%
\begin{gather*}
\delta _{+1/2}Q(x^{+})  = -\epsilon _{-}H(x^{+}), \\
\delta _{+3/2}Q(x^{+})  = \epsilon _{-}\left( \frac{1}{4}\left(
\partial _{+}\phi \right) ^{2}-\frac{1}{2}\psi _{+}\partial _{+}\psi
_{+}+\partial
_{+}\phi \psi _{+}Q(x^{+})-\frac{1}{4}H^{2}(x^{+})\right), \\
\delta _{+3/2}H(x^{+})  = \epsilon _{-}\left( \partial _{+}\phi
\partial _{+}\psi _{+}-\frac{1}{2}\partial _{+}^{2}\phi \psi
_{+}+\frac{1}{2}\partial _{+}\phi \psi _{+}H(x^{+})-Q(x^{+})\partial
_{+}H(x^{+})\right).
\end{gather*}

Applying $\delta _{+3/2}$ twice we get a local f\/low description of
the hierarchy for $t_{+1}$ and $t_{+3}$ in terms of the sinh-Gordon
variables used to described it in terms of $t_{+1}$ and $t_{-1},$ cf.\
equation (\ref{t-1 equations})%
\begin{gather}
4\partial _{+3}\phi  = \partial _{+}^{3}\phi -2\left( \partial
_{+}\phi
\right) ^{3}-3\partial _{+}\phi \psi _{+}\partial _{+}\psi _{+},  \qquad
4\partial _{+3}\psi _{+}  = \partial _{+}^{3}\psi _{+}-3\partial
_{+}\phi
\partial _{+}\left( \partial _{+}\phi \psi _{+}\right),  \notag \\
4\partial _{+3}\psi _{-}  = \big( 2\left( \partial _{+}\phi \right)
^{2}\psi _{+}-\partial _{+}^{2}\psi _{+}\big) h(\phi ),
\label{(1,1) mKdV equations 1}
\end{gather}
where $\big[ \delta _{+3/2},\delta _{+3/2}^{\prime }\big] (\ast
)=-2\epsilon _{1}\epsilon _{2}\partial _{+3}(\ast )$ in agreement with $\big\{ F_{1}^{(+3/2)},F_{1}^{(+3/2)}\big\} =-2E_{+}^{(+3)}$. Introducing $u=\partial _{+}\phi $ (the space variable $x$ is described by
$t_{+1}$) we recover the super mKdV equations
\begin{gather*}
4\partial _{+3}u  = u^{\prime \prime \prime }-6u^{2}u^{\prime
}-3\psi
_{+}\left( u\psi _{+}^{\prime }\right) ^{\prime }, \qquad
4\partial _{+3}\psi _{+}  = \psi _{+}^{\prime \prime \prime
}-3u\left( u\psi
_{+}\right) ^{\prime }, \\
4\partial _{+3}\psi _{-}  = \left( 2u^{2}\psi _{+}-\psi _{+}^{\prime
\prime }\right) h(\partial ^{-1}u).
\end{gather*}
The $\psi _{-}$ equation is non-local and is remnant from the
negative part of the hierarchy. The supersymmetry has to be reduced
to the usual $N=(1,0)$
in order to have a local description in terms of the mKdV variables $u$ and $\psi _{+}$. Note that $Q(x^{+})$ and $H(x^{+})$ resembles a piece of
the supercharge and a component of the stress tensor.

\begin{remark}
The higher grade fermionic transformation are some sort of `square
root' of the isospectral f\/lows, this can be seen also in
\cite{Mathieu} where Poisson brackets were used (compare with~(\ref{flows})).
\end{remark}

The positive and negative parts of the extended mKdV hierarchy
carries exactly the same information when considered separately and
to obtain relativistic equations we combine their Lax pairs in
dif\/ferent gauges as explained above by general arguments. The
potential couples the two sectors and enters through the functional
$h(\phi )$. We check this explicitly by considering $\left(
t_{+1},t_{+3}\right) $, $\left( t_{-1},t_{-3}\right) $ and compute
$L_{\pm 3}$ in order to construct a 1-soliton solution which solves
any equation of the hierarchy mixing the four times $\left(
t_{+1},t_{+3},t_{-1},t_{-3}\right) $. We keep the name soliton
solution for simplicity but strictly speaking we need to have a
multi-vacuum theory in order to def\/ine the asymptotic boundary
values of the f\/ield solution.

The algebraic dressing technique (recall equation (\ref{Isospectral
evolutions})
and (\ref{Isospectral II})) suggest the following forms for the Lax operators%
\begin{gather}
L_{+3}  = \partial
_{+3}+D^{(0)}+D^{(+1/2)}+D^{(+1)}+D^{(+3/2)}+D^{(+2)}+D^{(+5/2)}+E^{(+3)},
\notag \\
L_{-3}  = \partial _{-3}-B\big(
D^{(-1/2)}+D^{(-1)}+D^{(-3/2)}+D^{(-2)}+D^{(-5/2)}+E^{(-3)}\big)
B^{-1}. \label{L-3}
\end{gather}%
From $\left[ L_{+},L_{+3}\right] =0$ we get for the positive part of
the
hierarchy, following \cite{half-integer gradations}, the solution
\begin{gather*}
D^{(+5/2)}  = (-2\psi _{+}) G_{1}^{(+5/2)}, \qquad
D^{(+2)}   = (-\partial _{+}\phi )M_{1}^{(+2)}, \\
D^{(+3/2)}  = \left( -\psi _{+}\partial _{+}\phi \right)
F_{1}^{(+3/2)}+\left( \partial _{+}\psi _{+}\right) G_{2}^{(+3/2)}, \\
D^{(+1)} =\left( -2\psi _{+}\partial _{+}\psi _{+}\right)
K_{1}^{(+1)}+\left( \frac{1}{2}\left( \partial _{+}\phi \right)
^{2}+2\psi _{+}\partial _{+}\psi _{+}\right) K_{2}^{(+1)}+\left(
\frac{1}{2}\partial
_{+}^{2}\phi \right) M_{2}^{(+1)}, \\
D^{(+1/2)} =-\frac{1}{2}\left( \partial _{+}^{2}\phi \psi
_{+}-\partial
_{+}\phi \partial _{+}\psi _{+}\right) F_{2}^{(+1/2)}+\left( -\frac{1}{2}%
\partial _{+}^{2}\psi _{+}+\psi _{+}\left( \partial _{+}\phi \right)
^{2}\right) G_{1}^{(+1/2)}, \\
D^{(0)} =\left( -\frac{1}{4}\partial _{+}^{3}\phi
+\frac{1}{2}\left(
\partial _{+}\phi \right) ^{3}+3\partial _{+}\phi \psi _{+}\partial _{+}\psi
_{+}\right) M_{1}^{(0)}.
\end{gather*}%
The equation of motion are given by the degree zero component and
are, as expected, given the f\/irst two equations of (\ref{(1,1) mKdV
equations 1}), after taking $\psi _{+}\rightarrow \frac{1}{2}\psi
_{+}$. Now, performing a gauge transformation with $B^{-1}$ in order
to eliminate the $B$ conjugation on $L_{-3}$ and $L_{-}$ we change
to
\begin{gather*}
L_{-}  = \partial _{-}+B^{-1}\partial _{-}B-\big( E_{-}^{(-1)}+\psi
_{-}^{\left( -1/2\right) }\big), \\
L_{-3}  = \partial _{-3}+B^{-1}\partial _{-3}B-\big(
D^{(-1/2)}+D^{(-1)}+D^{(-3/2)}+D^{(-2)}+D^{(-5/2)}+E^{(-3)}\big)
\end{gather*}%
and from $\left[ L_{-},L_{-3}\right] =0$ we get for the negative
part of the
hierarchy, the solutions%
\begin{gather*}
D^{(-5/2)}  = \left( 2\psi _{-}\right) G_{2}^{(-5/2)}, \qquad
D^{(-2)}  = \left( -\partial _{-}\phi \right) M_{1}^{(-2)}, \\
D^{(-3/2)}  = \left( \psi _{-}\partial _{-}\phi \right)
F_{2}^{(-3/2)}+\left( \partial _{-}\psi _{-}\right) G_{1}^{(-3/2)}, \\
D^{(-1)}  = \left( -2\psi _{-}\partial _{-}\psi _{-}\right)
K_{1}^{(-1)}+\left( \frac{1}{2}\left( \partial _{-}\phi \right)
^{2}+2\psi _{-}\partial _{-}\psi _{-}\right) K_{2}^{(-1)}+\left(
-\frac{1}{2}\partial
_{-}^{2}\phi \right) M_{2}^{(-1)}, \\
D^{(-1/2)}  = -\frac{1}{2}\left( \partial _{-}^{2}\phi \psi
_{-}-\partial
_{-}\phi \partial _{-}\psi _{-}\right) F_{1}^{(-1/2)}+\left( \frac{1}{2}%
\partial _{-}^{2}\psi _{-}-\left( \partial _{-}\phi \right) ^{2}\psi
_{-}\right) G_{2}^{(-1/2)}.
\end{gather*}%
The equations of motion are now%
\begin{gather*}
4\partial _{-3}\phi  = \partial _{-}^{3}\phi -2\left( \partial
_{-}\phi
\right) ^{3}-3\partial _{-}\phi \psi _{-}\partial _{-}\psi _{-}, \qquad
4\partial _{-3}\psi _{-}  = \partial _{-}^{3}\psi _{-}-3\partial
_{-}\phi
\partial _{-}\left( \partial _{-}\phi \psi _{-}\right)
\end{gather*}
after taking $\psi _{-}\rightarrow \frac{1}{2}\psi _{-}$. With
$v=\partial
_{-}\phi $ (the space variable $x$ is described by $t_{-1}$) we get
\begin{gather*}
4\partial _{-3}v  = v^{\prime \prime \prime }-6v^{2}v-3\psi
_{-}\left( v\psi
_{-}^{\prime }\right) ^{\prime }, \qquad
4\partial _{-3}\psi _{-} = \psi _{-}^{\prime \prime \prime
}-3v\left( v\psi _{-}\right) ^{\prime }.
\end{gather*}
From the solution of the equation (\ref{-3 equation}), i.e.\
$D^{(0)}=-\partial _{+3}BB^{-1}$ we conf\/irm that $4\partial
_{+3}\phi =\partial _{+}^{3}\phi -2\left( \partial _{+}\phi \right)
^{3}-3\partial _{+}\phi \psi _{+}\partial _{+}\psi _{+}$, after
taking $\psi _{+}\rightarrow \frac{1}{2}\psi _{+}$. This is the
$N=(0,1)$ mKdV equation with opposite chirality.

The simplest example of a relativistic equation is provided by the
sinh-Gordon model which corresponds to the lowest $t_{\pm 1}$
times. Now
compute $\left[ L_{+3},L_{-3}\right] =0$ with $L_{-3}$ in the form (\ref{L-3}), i.e.\ in its original gauge because the two copies are to be taken
in dif\/ferent gauges. The generalized relativistic system of
equations is given in the Appendix~\ref{appendixB}, equation~(\ref{higher eq of
motion}), and is a generalization to the $t_{\pm 3}$ times. The two
sectors of the extended hierarchy are identical, thus we complete
the system (\ref{(1,1) mKdV
equations 1}) with the following set of equations
\begin{gather}
4\partial _{-3}\phi  = \partial _{-}^{3}\phi -2\left( \partial
_{-}\phi \right) ^{3}-3\partial _{-}\phi \psi _{-}\partial _{-}\psi
_{-},
\qquad
4\partial _{-3}\psi _{-}  = \partial _{-}^{3}\psi _{-}-3\partial
_{-}\phi
\partial _{-}\left( \partial _{-}\phi \psi _{-}\right),  \notag \\
4\partial _{-3}\psi _{+}  = \big( 2\left( \partial _{-}\phi \right)
^{2}\psi _{-}-\partial _{-}^{2}\psi _{-}\big) h(\phi )  \label{(1,1) mKdV equation 2}
\end{gather}
and due to this symmetric behavior we can easily read of\/f the
$\delta _{-3/2} $ transformations directly from (\ref{3/2 variations}) simply by replacing all $+$ sub-indexes by $-$
sub-indexes.

Now consider for simplicity the bosonic limit of the generalized equations (\ref{higher eq of motion}). It is given by
\begin{gather}
\partial _{+3}\partial _{-3}\phi  = -2\sinh [2\phi ]+\frac{1}{2}\big(
\left( \partial _{+}\phi \right) ^{2}\left( \partial _{-}\phi
\right) ^{2}+\partial _{+}^{2}\phi \partial _{-}^{2}\phi \big)
\sinh [2\phi ]
\notag \\
\phantom{\partial _{+3}\partial _{-3}\phi  =}{}
-\frac{1}{2}\big( \partial _{+}^{2}\phi \left( \partial _{-}\phi
\right)
^{2}+\left( \partial _{+}\phi \right) ^{2}\partial _{-}^{2}\phi
\big) \cosh [2\phi ],  \notag \\
\partial _{\pm 3}(\partial _{\mp }\phi )   = \left( \partial _{\pm
}\phi \right) ^{2}\sinh [2\phi ]-\partial _{\pm }^{2}\phi \cosh
[2\phi ],
\notag \\
\partial _{\pm 3}\left( \partial _{\mp }\phi \right) ^{2}  = 2\partial _{\mp
}\phi \left( \partial _{\pm }\phi \right) ^{2}\sinh [2\phi
]-2\partial _{\mp
}\phi \partial _{\pm }^{2}\phi \cosh [2\phi ],  \notag \\
\partial _{\pm 3}\left( \partial _{\mp }^{2}\phi \right)  = 4\partial _{\pm
}\phi +2\partial _{\mp }\phi \left( \partial _{\pm }\phi \right)
^{2}\cosh [2\phi ]-2\partial _{\mp }\phi \partial _{\pm }^{2}\phi
\sinh [2\phi ] \label{higher sinh-gordon equation}
\end{gather}
and describe the behavior of the `$\phi $-descendants' $\partial _{\pm }\phi,\left( \partial _{\pm
}\phi \right) ^{2}$ and $\partial _{\pm }^{2}\phi $ in terms of
opposite $t_{\mp 3} $ times. The equation (\ref{higher sinh-gordon equation}) can be obtained alternatively by using the basic
non-linear relations $\partial _{\pm 3}\phi =\frac{1}{4}\partial
_{\pm }^{3}\phi -\frac{1}{2}\left(
\partial _{\pm }\phi \right) ^{3}$, and the lowest relativistic equation $%
\partial _{+}\partial _{-}\phi =-2\sinh [2\phi ].$ The whole set of
equations associated to the times $\left(
t_{-3},t_{-1},t_{+1},t_{+3}\right) $ is completed by the equations
extracted from the relations $\left[ L_{+},L_{-3}\right] =0$ and
$\left[ L_{-},L_{+3}\right] =0.$ They are given
by the sinh-Gordon equation itself, the equations on the third line of (\ref{higher sinh-gordon equation}) and
\begin{gather}
4\partial _{\mp }\phi  = \partial _{\pm }\left( \partial _{\mp }\phi
\right) ^{2}\sinh [2\phi ]-\partial _{\pm }\partial _{\mp }^{2}\phi
\cosh [2\phi ],
 \notag \\
0  = \partial _{\pm }\partial _{\mp }^{2}\phi \sinh [2\phi
]-\partial _{\pm }\left( \partial _{\mp }\phi \right) ^{2}\cosh
[2\phi ]. \label{mixed equation}
\end{gather}

Now take $K^{\pm (n)}=E^{(\pm n)}$ in (\ref{non-abelian flows}) and
conjugate them with the grading operator $Q$ as follows%
\begin{gather*}
E^{\prime (\pm n)}=\exp (\alpha Q)E^{(\pm n)}\exp (-\alpha Q)=\exp
(\pm n)E^{(\pm n)}=\Lambda ^{\pm n}E^{(\pm n)},
\end{gather*}
where $\Lambda =\exp (\alpha ).$ The equations are invariant under
these rescalings and the Lorentz transformations, i.e.\ $x^{\prime \pm
}=\Lambda
^{\pm 1}x^{\pm }$ can be generalized to the whole set of f\/lows by taking $%
t_{\pm n}^{\prime }=\Lambda ^{\pm n}t_{\pm n}$.

To f\/ind the 1-soliton solution of the equation (\ref{higher sinh-gordon equation}) we use the dressing method which is another
application of the formalism given above in Section~\ref{section2.1}, see also
\cite{Guilherme} for
computational details. The four vacuum Lax operators involved are%
\begin{gather*}
L_{\pm }=\partial _{\pm }\pm E_{\pm }^{(\pm 1)},\qquad L_{\pm
3}=\partial _{\pm 3}\pm E^{(\pm 3)}
\end{gather*}
and the zero curvature conditions imply they are pure gauge $A_{i}^{V}=T_{0}^{-1}\partial _{i}T_{0}$. Hence we have as usual
\begin{gather*}
A_{+}^{V}  = E_{+}^{(+1)}-x^{-}C,\qquad A_{-}^{V}=-E_{-}^{(-1)}, \qquad
A_{+3}^{V}  = E^{(+3)}-3t_{-3}C,\qquad A_{-3}^{V}=-E^{(-3)}, \\
T_{0}  = \exp \big( t_{+3}E^{(+3)}+x^{+}E_{+}^{(+1)}\big) \exp
\big( -x^{-}E_{-}^{(-1)}-t_{-3}E^{(-3)}\big),
\end{gather*}
where we have used $\left[ E^{(m)},E^{(n)}\right]
=\frac{1}{2}(m-n)\delta _{m+n,0}C$ for $(m,n)$ odd integers. The
dressing of a vacuum Lax connections $A_{i}^{V}$ is the gauge
transformation $A_{i}=\left( \Theta _{\pm }\right)
^{-1}A_{i}^{V}\Theta _{\pm }+\left( \Theta _{\pm }\right)
^{-1}\partial _{i}\Theta _{\pm },$ satisfying $\Theta _{-}\Theta
_{+}^{-1}=T_{0}^{-1}gT_{0},$ where $g$ is an arbitrary constant
group element. Assuming that $\Theta
_{-}^{-1}=e^{p(-1)}e^{p(-2)}\cdots $, $\Theta
_{+}^{-1}=e^{q(0)}e^{q(1)}e^{q(2)}\cdots$, where $p(-i)$ and $q(i)$ are
linear combinations of grade $(-i)$ and $(+i)$ respectively, the
zero grade
component of $A_{i}$ leads to the solution%
\begin{gather*}
e^{q(0)}=B^{-1}e^{-\nu C},\qquad B=\exp \left( \phi H\right).
\end{gather*}%
From this we have%
\begin{equation*}
\big\langle \lambda ^{\prime }\big\vert B^{-1}\big\vert \lambda
\big\rangle e^{-\nu }=\big\langle \lambda ^{\prime }\big\vert
T_{0}^{-1}gT_{0}\big\vert \lambda \big\rangle,
\end{equation*}%
where $|\lambda \rangle$ and $\langle \lambda ^{\prime }|$ are
annihilated
by the grade $(-i)$ and $(i)$ generators. Taking the highest weight states $
|\lambda _{i}\rangle $, $i=0,1$ of $sl(2)^{(1)}$ we get the following tau-functions
\begin{gather*}
\tau _{1}  = e^{-\phi -\nu }=\big\langle \lambda _{1}\big\vert
T_{0}^{-1}gT_{0}\big\vert \lambda _{1}\big\rangle,\qquad
\tau _{0}  = e^{-\nu }=\big\langle \lambda _{0}\big\vert
T_{0}^{-1}gT_{0}\big\vert \lambda _{0}\big\rangle.
\end{gather*}

The so called solitonic specialization corresponds to the situation
when $g$ is given by the exponential of an eigenvalue $F(z)$, $z\in
\mathbb{C}
$ of the operator ${\rm ad}\, E^{(k)}.$ In this case we have%
\begin{equation*}
\big[ E^{(n)},F(z)\big] =-2z^{n}F(z),
\end{equation*}
where $F(z)$ is the vertex operator%
\begin{equation*}
F(z)=\sum\limits_{n=-\infty }^{+\infty }\left[ \left( M_{1}^{(2n)}-\frac{1}{2}\delta _{n,0}C\right) z^{-2n}+M_{2}^{(2n+1)}z^{-2n-1}\right].
\end{equation*}%
With the following result
\begin{gather}
T_{0}^{-1}gT_{0}  = \exp \left\{ \rho (z)F(z)\right\},  \qquad
\rho (\gamma )  = \exp 2\left\{
-t_{-3}z^{-3}-x^{-}z^{-1}+x^{+}z+t_{+3}z^{3}\right\},  \label{rho}
\end{gather}
we get the 1-soliton solution depending on the f\/irst four times of
the hierarchy
\begin{equation}
\phi (t_{-3},x^{-},x^{+},t_{+3})=-\log \left( \frac{\tau _{1}}{\tau _{0}}
\right),  \label{soliton}
\end{equation}%
where $\tau _{1}=1+\frac{b}{2}\rho (z)$, $\tau _{0}=
1-\frac{b}{2}\rho (z)$ and $b={\rm const}$. The f\/ield~(\ref{soliton}) is a
simultaneous solution of the
bosonic limits (with $\mu =1$) of the equations~(\ref{t-1 equations}), (\ref{(1,1) mKdV equations 1}), (\ref{(1,1) mKdV equation 2}) and the
whole set of equations~(\ref{higher sinh-gordon equation}) and~(\ref{mixed equation}).
The interesting point is that the second and third terms of the RHS of~(\ref{higher sinh-gordon equation}) cancel each other and the f\/ield $\phi
$ has
to obey
\begin{equation*}
\partial _{+3}\partial _{-3}\phi =-2\sinh [2\phi ],
\end{equation*}
which is the case. Then, the solution (\ref{soliton}) behaves under
$t_{\pm 3}$ like a sinh-Gordon soliton solution but with a cubic
rapidity. For higher grade times the extension in~(\ref{rho}) is
direct. At this point we can notice that each chirality of the
extended hierarchy is attached separately to the poles $z=+\infty $
(positive f\/lows) and $z=0$ (negative
f\/lows) of the Lax pair with spectral parameter~$z$, see for instance~\cite{BBT}. This is exactly the pole structure entering the def\/inition
of the sinh-Gordon Lax pair.

We end by giving the fermionic currents for the times $t_{\pm 3}$ and $D^{(+1/2)}$. After a lengthly but straightforward computation, the equation (\ref{generalized conservation laws}) becomes $\partial
_{+3}J_{-3}^{(+1/2)}+\partial _{-3}J_{+3}^{(+1/2)}=0$ with the
current components
\begin{gather*}
-\frac{1}{8}j_{+3}^{(+1/2)}  = \partial _{-}\phi \partial _{-}\psi
_{-}-\psi
_{-} \partial _{-}^{2}\phi, \\
-\frac{1}{8}j_{-3}^{(+1/2)}  = -2\psi _{+}\sinh [\phi ]-\left( \frac{1}{2}%
\partial _{+}^{2}\psi _{+}\partial _{+}\phi -\psi _{+}\left( \partial
_{+}\phi \right) ^{3}-\frac{1}{2}\partial _{+}\psi _{+}\partial
_{+}^{2}\phi \right) \cosh [\phi ].
\end{gather*}

\subsection[The $N=(2,2)$ Landau-Ginzburg Toda model]{The $\boldsymbol{N=(2,2)}$ Landau--Ginzburg Toda model}
\label{section3.2}

Take the $psl(2,2)_{\left[ 3\right] }^{(2)}$ superalgebra (see
Appendix~\ref{appendixA} for details). The Lagrangian is given by
\begin{gather*}
L  = -\frac{k}{2\pi }\left[ \partial _{+}\phi _{1}\partial _{-}\phi
_{1}+\partial _{+}\phi _{3}\partial _{-}\phi _{3}+\psi _{1}\partial
_{+}\psi
_{1}+\psi _{3}\partial _{+}\psi _{3}+\overline{\psi }_{2}\partial _{-}
\overline{\psi }_{2}+\overline{\psi }_{4}\partial _{-}\overline{\psi }_{4}-V
\right], \\
V  = 2\cosh [2\phi _{1}]-2\cos [2\phi _{3}]-4\left( \psi _{1}\overline{\psi }%
_{2}+\psi _{3}\overline{\psi }_{4}\right) \cosh [\phi _{1}]\cos
[\phi _{3}]
\\
\phantom{V  =}{} +4\left( \psi _{1}\overline{\psi }_{4}-\psi _{3}\overline{\psi }
_{2}\right) \sinh [\phi _{1}]\sin [\phi _{3}].
\end{gather*}

Taking $D^{(+1/2)}=D_{1}^{(+1/2)}+D_{2}^{(+1/2)}$ with $D_{1}^{(+1/2)}=%
\epsilon _{1}F_{1}^{(+1/2)}$, \ $D_{2}^{(+1/2)}=\epsilon
_{2}F_{4}^{(+1/2)}$
and $D^{(-1/2)}=D_{1}^{(-1/2)}+D_{2}^{(-1/2)}$ with $D_{1}^{(-1/2)}=%
\overline{\epsilon }_{1}F_{2}^{(-1/2)},D_{2}^{(-1/2)}=\overline{\epsilon }%
_{2}F_{3}^{(-1/2)}$, the supersymmetry transformations (\ref{susy
transformations 1/2}) and (\ref{susy transformations -1/2}) with
$\psi _{i}$,
$\overline{\psi }_{i}\rightarrow $ $\frac{1}{2}\psi _{i}$, $\frac{1}{2}\overline{\psi }_{i}$ are
\begin{gather}
\delta _{+1/2}\phi _{1}  = +2\left( \epsilon _{1}\overline{\psi }%
_{4}-\epsilon _{2}\overline{\psi }_{2}\right),\qquad
\delta _{+1/2}\phi _{3}=-2\left( \epsilon _{2}\overline{\psi
}_{4}+\epsilon
_{1}\overline{\psi }_{2}\right),  \notag \\
\delta _{+1/2}\overline{\psi }_{2}  = +\left( \epsilon _{2}\partial
_{+}\phi _{1}+\epsilon _{1}\partial _{+}\phi _{3}\right),\qquad
\delta _{+1/2}\overline{\psi }_{4}=-\left( \epsilon
_{1}\partial _{+}\phi
_{1}-\epsilon _{2}\partial _{+}\phi _{3}\right),  \notag \\
\delta _{+1/2}\psi _{1}  = -2\epsilon _{1}\cosh [\phi _{1}]\sin
[\phi
_{3}]-2\epsilon _{2}\sinh [\phi _{1}]\cos [\phi _{3}],  \notag \\
\delta _{+1/2}\psi _{3}  = +2\epsilon _{1}\sinh [\phi _{1}]\cos
[\phi
_{3}]-2\epsilon _{2}\cosh [\phi _{1}]\sin [\phi _{3}],  \notag \\
\delta _{-1/2}\phi _{1}  = +2\left( \overline{\epsilon }_{1}\psi _{3}-%
\overline{\epsilon }_{2}\psi _{1}\right),\qquad
\delta _{-1/2}\phi _{3}=+2\left( \overline{\epsilon }_{1}\psi _{1}+\overline{%
\epsilon }_{2}\psi _{3}\right),  \notag \\
\delta _{-1/2}\psi _{1}  = +\left( \overline{\epsilon }_{2}\partial
_{-}\phi _{1}-\overline{\epsilon }_{1}\partial _{-}\phi _{3}\right)
,\qquad \delta _{-1/2}\psi _{3}=-\left(
\overline{\epsilon }_{1}\partial _{-}\phi
_{1}+\overline{\epsilon }_{2}\partial _{-}\phi _{3}\right),  \notag \\
\delta _{-1/2}\overline{\psi }_{2}  = -2\overline{\epsilon
}_{1}\cosh [\phi _{1}]\sin [\phi _{3}]+2\overline{\epsilon
}_{2}\sinh [\phi _{1}]\cos [\phi
_{3}],  \notag \\
\delta _{-1/2}\overline{\psi }_{4}  = -2\overline{\epsilon
}_{1}\sinh [\phi _{1}]\cos [\phi _{3}]-2\overline{\epsilon
}_{2}\cosh [\phi _{1}]\sin [\phi _{3}]. \label{(2,2) transformations}
\end{gather}

We can check (\ref{homo}) by applying (\ref{(2,2) transformations})
twice giving
\begin{gather*}
\big[ \delta _{+1/2},\delta'
_{+1/2}\big]  = +4\left( \epsilon _{1}\epsilon _{1}^{\prime
}+\epsilon
_{2}\epsilon _{2}^{\prime }\right) \partial _{+}, \qquad
\big[ \delta _{-1/2},\delta'
_{-1/2}\big]  = +4\left( \overline{\epsilon }_{1}\overline{\epsilon }
_{1}^{\prime }+\overline{\epsilon }_{2}\overline{\epsilon
}_{2}^{\prime
}\right) \partial _{-}, \\
\big[ \delta _{+1/2},\delta _{-1/2}\big]  = 0.
\end{gather*}

We have four real supercharges $N=(2,2)$ because
$\dim\mathcal{K}_{F}^{(\pm 1/2)}=2$. They are extracted from~(\ref{Q+}) and (\ref{Q-}) and are given by

f\/low $\delta _{+1/2}$:
\begin{gather*}
 Q^{\left( -1/2\right)
}=Q_{+}^{1}F_{2}^{(-1/2)}+Q_{+}^{2}F_{3}^{(-1/2)}, \\
Q_{+}^{1}  = \int dx^{1}\left( \overline{\psi }_{2}\partial _{+}\phi _{3}-%
\overline{\psi }_{4}\partial _{+}\phi _{1}-2\psi _{1}\cosh [\phi
_{1}]\sin
[\phi _{3}]+2\psi _{3}\sinh [\phi _{1}]\cos [\phi _{3}]\right), \\
Q_{+}^{2} = \int dx^{1}\left( \overline{\psi }_{4}\partial _{+}\phi _{3}+%
\overline{\psi }_{2}\partial _{+}\phi _{1}-2\psi _{1}\sinh [\phi
_{1}]\cos
[\phi _{3}]-2\psi _{3}\cosh [\phi _{1}]\sin [\phi _{3}]\right),
\end{gather*}

f\/low $\delta _{-1/2}$:
\begin{gather*}
Q^{\left( +1/2\right)
}=Q_{-}^{1}F_{1}^{(+1/2)}+Q_{-}^{2}F_{4}^{(+1/2)}, \\
Q_{-}^{1}  = \int dx^{1}\left( \psi _{1}\partial _{-}\phi _{3}+\psi
_{3}\partial _{-}\phi _{1}+2\overline{\psi }_{2}\cosh [\phi
_{1}]\sin [\phi
_{3}]+2\overline{\psi }_{4}\sinh [\phi _{1}]\cos [\phi _{3}]\right), \\
Q_{-}^{2}  = \int dx^{1}\left( \psi _{3}\partial _{-}\phi _{3}-\psi
_{1}\partial _{-}\phi _{1}-2\overline{\psi }_{2}\sinh [\phi
_{1}]\cos [\phi _{3}]+2\overline{\psi }_{4}\cosh [\phi _{1}]\sin
[\phi _{3}]\right).
\end{gather*}

Now introduce the complex f\/ields
\begin{equation*}
\psi _{-}=\psi _{1}+i\psi _{3},\qquad \psi _{+}=-\overline{\psi }_{2}+i
\overline{\psi }_{4},\qquad \phi =\phi _{3}+i\phi _{1}
\end{equation*}%
and the superpotential $W\left( \phi \right) =2\mu \cos \phi $ in
order to
write the lagrangian in a more familiar form. Then, we have
\begin{gather}
L  = -\frac{k}{2\pi }\left[ \partial _{+}\phi \partial _{-}\phi
^{\ast }+\psi _{-}^{\ast }\partial _{+}\psi _{-}+\psi _{+}^{\ast
}\partial _{-}\psi
_{+}-V\right],  \notag \\
V  = \left\vert W^{\prime }\left( \phi \right) \right\vert
^{2}-\left[ W^{\prime \prime }\left( \phi \right) \psi _{-}\psi
_{+}+W^{\ast \prime \prime }\left( \phi ^{\ast }\right) \psi
_{-}^{\ast }\psi _{+}^{\ast }\right],  \label{CY-Landau Ginzburg}
\end{gather}
where $\left\vert W^{\prime }\left( \phi \right) \right\vert
^{2}=2\mu ^{2}\cosh [2\phi _{1}]-2\mu ^{2}\cos [2\phi _{3}].$ This
Lagrangian is invariant under the common $N=(2,2)$ superspace
transformations for a complex chiral bosonic superf\/ield. In terms of
the new complex f\/ields we have for (\ref{(2,2) transformations})
with $\epsilon _{-}=-(\epsilon
_{1}+i\epsilon _{2})$ and $\epsilon _{+}=\overline{\epsilon }_{1}-i\overline{\epsilon }_{2}$ that%
\begin{alignat*}{3}
& \delta _{+1/2}\phi  = -2\epsilon _{-}\psi _{+},\qquad && \delta _{-1/2}\phi =+2\epsilon _{+}\psi _{-},&
 \notag \\
& \delta _{+1/2}\psi _{+}  = +\epsilon _{-}^{\ast }\partial _{+}\phi
,\qquad && \delta _{-1/2}\psi
_{+}=-i\epsilon _{+}W^{\prime
\ast }\left( \phi ^{\ast }\right),&  \notag \\
& \delta _{+1/2}\psi _{-}  = -i\epsilon _{-}W^{\prime \ast }\left(
\phi ^{\ast }\right),\qquad && \delta _{-1/2}\psi
_{-}=-\epsilon _{+}^{\ast }\partial _{-}\phi & %\label{2,2 in complex variables}
\end{alignat*}
plus their complex conjugates. Def\/ine now the following complex
combinations of the supercharges $\frac{1}{2}Q_{\pm }=\mp Q_{\pm
}^{1}+iQ_{\pm }^{2}$
\begin{gather*}
Q_{\pm }  = 2\int dx^{1}\left( \psi _{\pm }\partial _{\pm }\phi
^{\ast }\mp
i\psi _{\mp }^{\ast }W^{\prime \ast }\left( \phi ^{\ast }\right) \right),\qquad
\overline{Q}_{\pm }  = 2\int dx^{1}\left( \psi _{\pm }^{\ast
}\partial _{\pm }\phi \pm i\psi _{\mp }W^{\prime }\left( \phi
\right) \right).
\end{gather*}
Now, with the Dirac brackets $\left\{ \phi,\partial _{t}\phi ^{\ast
}\right\} =\left\{ \phi ^{\ast },\partial _{t}\phi \right\} =+1$ and $%
\left\{ \psi _{+},\psi _{+}^{\ast }\right\} =\left\{ \psi _{-},\psi
_{-}^{\ast }\right\} =+1/2$ we have with $Q_{\pm }f=\left\{ Q_{\pm
},f\right\} $%
\begin{alignat*}{3}
& Q_{+}\phi = -2\psi _{+},\qquad && \overline{Q}_{+}\phi =0, &
\\
& Q_{+}\psi _{+} =0,\qquad && \overline{Q}%
_{+}\psi _{+}=\partial _{+}\phi, &  \\
& Q_{+}\psi _{-} = -iW^{\prime \ast }\left( \phi ^{\ast }\right)
,\qquad && \overline{Q}_{+}\psi _{-}=0, & \\
& Q_{-}\phi = -2\psi _{-},\qquad && \overline{Q}_{-}\phi =0, & \\
& Q_{-}\psi _{+}  = iW^{\prime \ast }\left( \phi ^{\ast }\right),\qquad && \overline{Q}_{-}\psi _{+}=0, &  \\
& Q_{-}\psi _{-}  = 0, \qquad && \overline{Q}_{-}\psi _{-}=\partial _{-}\phi &
\end{alignat*}
plus their complex conjugates.

Finally, the total variation becomes $\delta _{\rm SUSY}=\epsilon
_{-}Q_{+}-\epsilon _{+}Q_{-}+\epsilon _{-}^{\ast
}\overline{Q}_{+}-\epsilon
_{+}^{\ast } \overline{Q}_{-}$ with%
\begin{alignat*}{3}
& \delta _{\rm SUSY}\phi  = +2\epsilon _{+}\psi _{-}-2\epsilon _{-}\psi
_{+}, \qquad && \delta _{\rm SUSY}\phi ^{\ast }=+2\epsilon
_{+}^{\ast }\ \psi _{-}^{\ast }-2\epsilon _{-}^{\ast }\psi _{+}^{\ast }, &  \\
& \delta _{\rm SUSY}\psi _{+}  = +\epsilon _{-}^{\ast }\partial _{+}\phi
-i\epsilon _{+}W^{\prime \ast }\left( \phi ^{\ast }\right), \qquad && \delta _{\rm SUSY}\psi _{+}^{\ast }=+\epsilon _{-}\partial
_{+}\phi ^{\ast
}+i\epsilon _{+}^{\ast }\ W^{\prime }\left( \phi \right), &  \\
& \delta _{\rm SUSY}\psi _{-}  = -\epsilon _{+}^{\ast } \partial _{-}\phi
-i\epsilon _{-}W^{\prime \ast }\left( \phi ^{\ast }\right),\qquad && \delta _{\rm SUSY}\psi _{-}^{\ast }=-\epsilon _{+}\partial
_{-}\phi ^{\ast }+i\epsilon _{-}^{\ast }W^{\prime }\left( \phi
\right), &
\end{alignat*}
which are the usual $N=(2,2)$ supersymmetry transformations obtained
by using a superspace approach. As in the case of the $N=(1,1)$
model, we expect the existence of higher non-local fermionic f\/lows
for this model.

\begin{remark}
The action (\ref{CY-Landau Ginzburg}) is a Landau--Ginzburg model on
a f\/lat
non-compact trivial Calabi--Yau manifold $X$, i.e.\ $X=\mathbb{C}$. As is well known this model is $B$-twistable. It would be
interesting to see the relation of its chiral ring and the chiral
ring of a topologically
twisted version of a~superstring on ${\rm AdS}^{2}\times S^{2}$ where the action (\ref{CY-Landau Ginzburg}) is extracted as a Pohlmeyer reduction~\cite{tseytlin}. This would be in principle, a simple way to test up to
what point the Pohlmeyer reduction can be understood as an
equivalence of quantum f\/ield theories, at least at the level of
ground states and by eliminating conformal anomaly issues. It would
be also interesting to trace back (if
possible) the role played by the extended conformal symmetries of $\mathcal{W}$-type in terms of sigma model variables.
\end{remark}

\begin{remark}
In the relativistic sector of the AKNS hierarchy which is associated
to the
Homogeneous gradation, the Toda potential has a symmetry, i.e.~$\mathcal{K}%
_{B}^{(0)}\neq \varnothing $. In this case further reduction of the
model can be performed, eliminating the f\/lat directions. This is
done by coupling to a~quadratic gauge f\/ield $A$ and is equivalent to
the introduction of singular metrics def\/ining the non-Abelian Toda
models also known as singular Toda models. The minimal coupling of
the fermions with the gauge f\/ield gives two terms proportional to
$Q_{\pm }^{(0)}$. After integration of $A$, the action will have a
potential term which is quartic in the fermions, roughly
of the form%
\begin{equation*}
V=\big\langle Q_{+}^{(0)}BQ_{-}^{(0)}B^{-1}\big\rangle \sim
R(B)\psi _{+}^{1}\psi _{+}^{2}\psi _{-}^{1}\psi _{-}^{2},
\end{equation*}%
which is possibly related to the curvature $R$ of the background
metric. This hierarchy is relevant from the point of view of
Pohlmeyer reductions where reductions of non-linear sigma models
inside this hierarchy are quite common~\cite{tseytlin}.
\end{remark}

\section{Concluding remarks}\label{section4}

By coupling two identical supersymmetric integrable hierarchies we
have shown that the usual notion of superspace/supersymmetry is
embedded and alternatively described by the symmetry algebra spanned
by the subset of f\/lows $(t_{-1},t_{-1/2},t_{+1/2},t_{+1})\subset
\left( t_{\pm 1/2},t_{\pm 1},t_{\pm 3/2},\dots\right)$. We have given
the explicit form of the supercharges generating these extended
$(N,N)$ supersymmetry f\/lows and also shown that the higher grade
fermionic f\/lows are inevitably non-local on both chiral sectors. In
particular, when the $(t_{-1},t_{-1/2},t_{+1/2},t_{+1})$ f\/lows are
supplemented by the algebraic conditions $Q_{\pm }^{(0)}=0,$ the
integrable model is restricted to a reduced group manifold spanned
by the invariant subalgebra of a reductive automorphism $\tau
_{\rm red}$, see Appendix~\ref{appendixA} for an explicit example. We do not
supersymmetrize the f\/ields, i.e.\ the `angles' that parametrize the
group elements $\widehat{G}$ as usually done in the literature. The
reduction by $\tau _{\rm red}$ should be, in the general case, a natural
extension to superalgebras of the automorphism
used to def\/ine the bosonic af\/f\/ine (Abelian) Toda models, see \cite{top-antitop fusion} for an example applied to the Lie algebra $\widehat{\mathfrak{g}}=sl(m+1,\mathbb{C})$. The reduction provides a well def\/ined connection between the
dressing
elements and the physical degrees of freedom cf.\ equation~(\ref{current relation}), as well as the right number of terms in the potential appearing
in the action functional by truncating it at the second term, thus
the natural af\/f\/ine superalgebras involved in the supersymmetric
af\/f\/ine (Abelian) Toda models are the twisted ones.

What remains to be done in general terms is a formal proof of the
statement
that the introduction of $\tau _{\rm red}$ is responsible for the locality $(Q_{\pm }^{(0)}=0)$ of the lowest supersymmetry f\/lows $(\delta _{\pm
1/2}).$
This is equivalent to an explicit construction of a reductive automorphism $\tau _{\rm red}$ in which the invariant subalgebra $\widehat{\mathfrak{g}}_{\rm red}$ has no bosonic kernel of grade zero, i.e.\
$\mathcal{K}_{B}^{(0)}=\varnothing $. Such $\tau _{\rm red}$ will def\/ine in
principle, all supersymmetric af\/f\/ine Abelian Toda models attached to
the mKdV hierarchy (in the A series) and it would be interesting to
construct it also for other series of af\/f\/ine superalgebras. The next
step is to introduce a super tau-function formulation which we
expect to be a natural generalization of the one introduced in~\cite{Miramontes02} for the bosonic af\/f\/ine Toda mo\-dels, where an
inf\/inite number of conserved charges $Q_{n}^{\pm }$ were written in
terms
of the boundary va\-lues~$B$ of a single tau function $\tau $, in the form $Q_{n}^{\pm }\sim \partial _{\pm n}\log \tau |_{B}.$ As we are using
a fermionic version of the Toda models coupled to matter f\/ields
constructed in \cite{Ferreira}, it is possible that one has to
consider a `matrix' of tau functions $\tau _{mn}$ when solving the
whole system~(\ref{SUSY-Toda equations}). In this case, the
interpretation of the single $\tau $ function as a classical limit
of the partition function of some quantum integrable system will
change or will have to be modif\/ied in an appropriate way. The point
is that quantization can be done, in principle, by quantizing $\tau
_{mn}$, i.e.\ by using a quantum group of dressing transformations.
This obviously deserves a separate study.

%\looseness=1
A potential application for our supersymmetric af\/f\/ine Toda models is
related to the Pohlmeyer reduction of supersymmetric sigma models.
In the present construction we needed to impose the conditions
$Q_{\pm }^{(0)}=0$ in order to have $\mathcal{K}_{B}^{(0)}=\varnothing
$. This means that the corresponding supersymmetric reduced models
belong to the mKdV hierarchy and that no gauge
symmetries are involved, thus having Toda models of Abelian type (see equation~(\ref{CY-Landau Ginzburg}) for an example). To apply these results in
the most general situation we need to f\/ind a way of coupling two
supersymmetric AKNS hierarchies in which $\mathcal{K}_{B}^{(0)}\neq
\varnothing $, i.e.\ the reduced models will have gauge symmetries, thus
being Toda models of non-Abelian type. We expect to deduce the
action functional for the supersymmetric non-Abelian af\/f\/ine Toda
models similar to the one constructed in \cite[equation~(6.49)]{tseytlin} in the particular case of the reduction of the ${\rm AdS}_{5}\times
S^{5}$ superstring coset sigma model. We also expect to deduce the
supersymmetry transformations by treating them as fermionic symmetry
f\/lows in the AKNS hierarchy in the same way as it was done for the
action (\ref{SUSY-Toda action}) and (\ref{susy transformations
1/2}), (\ref{susy transformations -1/2}) in the mKdV hierarchy. It
is worth
to compare our supersymmetry transformations with the ones proposed in \cite[equations~(7.21), (7.22)]{tseytlin}. We also comment that our Lax
pair and action functional naturally includes a spectral parameter
and that it is also conformal invariant as a consequence of the
two-loop nature of the af\/f\/ine algebras used. The study of the `of\/f
shell' supersymmetric AKNS hierarchy and its symmetries is already
under investigation~\cite{yo2}.

\appendix

\section[Used superalgebras solving $Q_{\pm }^{(0)}=0$]{Used superalgebras solving $\boldsymbol{Q_{\pm }^{(0)}=0}$}\label{appendixA}

We consider the reductive automorphism $\tau _{\rm red}$ for
$sl(2,1)^{(1)}$
only \cite{half-integer gradations}. It is def\/ined by
\begin{alignat*}{3}
& \tau _{\rm red}\big( E_{\alpha }^{(n)}\big)  = -(-1)^{q\left(
E_{\alpha }^{(n)}\right) }E_{-\alpha }^{(n+\eta _{E_{\alpha
}})}\qquad && \text{for} \ \ \alpha
\in \text{bosonic root}, &  \\
& \tau _{\rm red}\big( E_{\alpha }^{(n)}\big)  = i(-1)^{q\left(
E_{\alpha }^{(n)}\right) }E_{-\alpha }^{(n+\eta _{E_{\alpha
}})}\qquad && \text{for} \ \ \alpha \in \text{fermionic root},&
\end{alignat*}%
where $q\big( E_{\alpha }^{(n)}\big) $ is the grade of $E_{\alpha
}^{(n)} $, i.e.\ $\big[ Q,E_{\alpha }^{(n)}\big] =q\big( E_{\alpha
}^{(n)}\big) E_{\alpha }^{(n)}$, $\eta _{E_{\alpha }}$ is def\/ined by $
\left[ \alpha \cdot H,E_{\alpha }\right] =\eta _{E_{\alpha }}E_{\alpha }$ and $
\alpha $ are the roots of the underlying f\/inite-dimensional Lie
superalgebra.

Invariance under $\tau _{\rm red}$ def\/ine a twisted superalgebra $sl(2,1)^{(2)}\subset sl(2,1)^{(1)}$. This twisted superalgebra solve
the
conditions $Q_{\pm }^{(0)}=0$. In~\cite{yo}, two sub-superalgebras of $sl(2,1)^{(1)}$ solving the conditions $Q_{\pm }^{(0)}=0$ where found
by
another method. The f\/irst algebra solution was denoted by $%
sl(2,1)_{[1]}^{(2)}$ and coincides with the subalgebra invariant
under $\tau
_{\rm red}$ above. The second algebra solution was denoted by $sl(2,1)_{[2]}^{(2)}.$ These two subalgebras gives rise to integrable
models of sinh-Gordon and sine-Gordon type which are related by
analytic continuation in the f\/ields. Similarly, in the case of
$psl(2,2)^{(1)}$, four sub-superalgebras solving the conditions
$Q_{\pm }^{(0)}=0$ were found. they were denoted as
$psl(2,2)_{[i]}^{(2)}$, $i=1,2,3,4$ and give rise to integrable
f\/ield theories coupling models of (sinh-sinh), (sine,sine),
(sinh,sine) and (sine, sinh) Gordon type respectively, all of them
related by analytic continuation. The subscript $[i]$ is just a
label used to denote a particular subalgebra solving the locallity
conditions $Q_{\pm }^{(0)}=0.$ The reason of using these subalgebras
is to turn local the lowest supersymmetry f\/lows. The explicit
dif\/ference, for the case $sl(2,1)^{(1)}$ (non-local) and
$sl(2,1)_{[1]}^{(2)}$ (local) was worked out in detail in~\cite{half-integer gradations}.

In this paper we consider, for the sake of simplicitly and with the
aim of
not being repetitive, only the superalgebras $sl(2,1)_{[1]}^{(2)}$ and $%
psl(2,2)_{[3]}^{(2)}$ whose def\/initions are as follows

$sl(2,1)_{[1]}^{(2)}$:
\begin{gather*}
Q=2d+\frac{1}{2}H_{1}, \qquad E_{\pm }^{(\pm 1)}=\lambda
_{2}\cdot H^{(\pm 1/2)}-\big( E_{\pm \alpha _{1}}^{(0)}+E_{\mp \alpha
_{1}}^{(\pm 1)}\big) \qquad \mbox{and}\\
\mathcal{K}_{B}=\big\{ K_{1}^{\left( 2n+1\right) }=\lambda
_{2}\cdot H^{\left( n+1/2\right) },\  K_{2}^{\left(
2n+1\right) }=E_{\alpha _{1}}^{\left( n\right) }+E_{-\alpha
_{1}}^{\left( n+1\right) }\big\},\\
\mathcal{M}_{B}=\big\{ M_{1}^{\left( 2n\right) }=H_{1}^{\left(
n\right) },\ M_{2}^{\left( 2n+1\right) }=E_{\alpha
_{1}}^{\left( n\right) }-E_{-\alpha _{1}}^{\left( n+1\right)
}\big\},\\
\mathcal{K}_{F}=\left\{
\begin{array}{c}
F_{1}^{\left( 2n+3/2\right) }=\big( E_{\alpha _{2}}^{\left(
n+1\right) }+E_{\alpha _{1}+\alpha _{2}}^{\left( n+1/2\right)
}\big) +\big( E_{-\alpha _{2}}^{\left( n+1/2\right) }+E_{-\alpha
_{1}-\alpha _{2}}^{\left(
n+1\right) }\big),  \vspace{2mm}\\
F_{2}^{\left( 2n+1/2\right) }=\big( E_{\alpha _{2}}^{\left(
n+1/2\right) }+E_{\alpha _{1}+\alpha _{2}}^{\left( n\right) }\big)
-\big( E_{-\alpha _{2}}^{\left( n\right) }+E_{-\alpha _{1}-\alpha
_{2}}^{\left( n+1/2\right) }\big)
\end{array}%
\right\},\\
\mathcal{M}_{F}=\left\{
\begin{array}{c}
G_{1}^{\left( 2n+1/2\right) }=\big( E_{\alpha _{2}}^{\left(
n+1/2\right) }-E_{\alpha _{1}+\alpha _{2}}^{\left( n\right)
}\big) +\big(
E_{-\alpha _{2}}^{\left( n\right) }-E_{-\alpha _{1}-\alpha _{2}
}^{\left( n+1/2\right) }\big),  \vspace{2mm}\\
G_{2}^{\left( 2n+3/2\right) }=\big( E_{\alpha _{2}}^{\left(
n+1\right) }-E_{\alpha _{1}+\alpha _{2}}^{\left(
n+1/2\right) }\big) -\big(
E_{-\alpha _{2}}^{\left( n+1/2\right) }-E_{-\alpha _{1}-\alpha _{2}
}^{\left( n+1\right) }\big)
\end{array}
\right\}.
\end{gather*}

$sl(2,2)_{[3]}^{(2)}$:
\begin{gather*}
Q=d+\frac{1}{2}\left( H_{1}+H_{3}\right), \qquad E_{\pm }^{(\pm
1)}=\big( E_{\pm \alpha _{1}}^{(0)}+E_{\mp \alpha _{1}}^{(\pm
1)}\big) +\big( E_{\pm \alpha _{3}}^{(0)}+E_{\mp \alpha
_{3}}^{(\pm 1)}\big) +I \qquad \mbox{and}\\
\mathcal{K}_{B}=\big\{ K_{1}^{\left( 2n+1\right) }=I^{\left( 2n+1\right) },
\  K_{2}^{\left( 2n+1\right) }=E_{\alpha _{1}}^{\left(
2n\right) }+E_{-\alpha _{1}}^{\left( 2n+2\right) }, \ K_{3}^{\left( 2n+1\right) }=E_{\alpha _{3}}^{\left( 2n+2\right)
}+E_{-\alpha _{3}}^{\left( 2n\right) }\big\},\\
\mathcal{M}_{B}=\left\{
\begin{array}{c}
M_{1}^{\left( 2n\right) }=H_{1}^{\left( 2n\right) },\ M_{2}^{\left( 2n+1\right) }=E_{\alpha _{1}}^{\left( 2n\right)
}-E_{-\alpha _{1}}^{\left(
2n+2\right) }, \vspace{2mm}\\
M_{3}^{\left( 2n+1\right) }=H_{3}^{\left( 2n+1\right) }, \
M_{4}^{\left( 2n\right) }=E_{\alpha _{3}}^{\left( 2n+1\right)
}-E_{-\alpha
_{3}}^{\left( 2n-1\right) }
\end{array}%
\right\},\\
\mathcal{K}_{F}=\left\{
\begin{array}{c}
F_{1}^{\left( 2n+1/2\right) }=\big( E_{\alpha _{2}}^{\left(
2n+1/2\right) }+E_{\alpha _{1}+\alpha _{2}+\alpha _{3}}^{\left(
2n+1/2\right) }\big) +\big( E_{\alpha _{1}+\alpha _{2}}^{\left(
2n-1/2\right) }+E_{\alpha
_{2}+\alpha _{3}}^{\left( 2n+3/2\right) }\big) + {} \vspace{1mm}\\
{}+\big( E_{-\alpha _{2}}^{\left( 2n+1/2\right) }+E_{-\alpha
_{1}-\alpha _{2}-\alpha _{3}}^{\left( 2n+1/2\right) }\big) +\big(
E_{-\alpha _{1}-\alpha _{2}}^{\left( 2n+3/2\right) }+E_{-\alpha
_{2}-\alpha
_{3}}^{\left( 2n-1/2\right) }\big),\vspace{2mm} \\
F_{2}^{\left( 2n+3/2\right) }=\big( E_{\alpha _{2}}^{\left(
2n+3/2\right) }+E_{\alpha _{1}+\alpha _{2}+\alpha _{3}}^{\left(
2n+3/2\right) }\big) +\big( E_{\alpha _{1}+\alpha _{2}}^{\left(
2n+1/2\right) }+E_{\alpha
_{2}+\alpha _{3}}^{\left( 2n+5/2\right) }\big) -{} \vspace{1mm}\\
{}-\big( E_{-\alpha _{2}}^{\left( 2n+3/2\right) }+E_{-\alpha
_{1}-\alpha _{2}-\alpha _{3}}^{\left( 2n+3/2\right) }\big) -\big(
E_{-\alpha _{1}-\alpha _{2}}^{\left( 2n+5/2\right) }+E_{-\alpha
_{2}-\alpha
_{3}}^{\left( 2n+1/2\right) }\big),\vspace{2mm}  \\
F_{3}^{\left( 2n+3/2\right) }=\big( E_{\alpha _{2}}^{\left(
2n+3/2\right) }+E_{\alpha _{1}+\alpha _{2}+\alpha _{3}}^{\left(
2n+3/2\right) }\big) -\big( E_{\alpha _{1}+\alpha _{2}}^{\left(
2n+1/2\right) }+E_{\alpha
_{2}+\alpha _{3}}^{\left( 2n+5/2\right) }\big) +{}\vspace{1mm} \\
{}+\big( E_{-\alpha _{2}}^{\left( 2n+3/2\right) }+E_{-\alpha
_{1}-\alpha _{2}-\alpha _{3}}^{\left( 2n+3/2\right) }\big) -\big(
E_{-\alpha _{1}-\alpha _{2}}^{\left( 2n+5/2\right) }+E_{-\alpha
_{2}-\alpha
_{3}}^{\left( 2n+1/2\right) }\big),\vspace{2mm}  \\
F_{4}^{\left( 2n+1/2\right) }=-\big( E_{\alpha _{2}}^{\left(
2n+1/2\right) }+E_{\alpha _{1}+\alpha _{2}+\alpha _{3}}^{\left(
2n+1/2\right) }\big) +\big( E_{\alpha _{1}+\alpha _{2}}^{\left(
2n-1/2\right) }+E_{\alpha
_{2}+\alpha _{3}}^{\left( 2n+3/2\right) }\big) +{} \vspace{1mm}\\
{}+\big( E_{-\alpha _{2}}^{\left( 2n+1/2\right) }+E_{-\alpha
_{1}-\alpha _{2}-\alpha _{3}}^{\left( 2n+1/2\right) }\big) -\big(
E_{-\alpha _{1}-\alpha _{2}}^{\left( 2n+3/2\right) }+E_{-\alpha
_{2}-\alpha _{3}}^{\left( 2n-1/2\right) }\big)
\end{array}%
\right\},\\
\mathcal{M}_{F}=\left\{
\begin{array}{c}
G_{1}^{\left( 2n+1/2\right) }=\big( E_{\alpha _{1}+\alpha
_{2}}^{\left( 2n-1/2\right) }-E_{\alpha _{2}+\alpha _{3}}^{\left(
2n+3/2\right) }\big) +\big( E_{\alpha _{2}}^{\left( 2n+1/2\right)
}-E_{\alpha _{1}+\alpha
_{2}+\alpha _{3}}^{\left( 2n+1/2\right) }\big) +{}\vspace{1mm} \\
{}+\big( E_{-\alpha _{1}-\alpha _{2}}^{\left( 2n+3/2\right)
}-E_{-\alpha _{2}-\alpha _{3}}^{\left( 2n-1/2\right) }\big)
+\big( E_{-\alpha _{2}}^{\left( 2n+1/2\right) }-E_{-\alpha
_{1}-\alpha _{2}-\alpha
_{3}}^{\left( 2n+1/2\right) }\big),\vspace{2mm} \\
G_{2}^{\left( 2n+3/2\right) }=\big( E_{\alpha _{1}+\alpha
_{2}}^{\left( 2n+1/2\right) }-E_{\alpha _{2}+\alpha _{3}}^{\left(
2n+5/2\right) }\big) +\big( E_{\alpha _{2}}^{\left( 2n+3/2\right)
}-E_{\alpha _{1}+\alpha
_{2}+\alpha _{3}}^{\left( 2n+3/2\right) }\big) -{}\vspace{1mm} \\
{}-\big( E_{-\alpha _{1}-\alpha _{2}}^{\left( 2n+5/2\right)
}-E_{-\alpha _{2}-\alpha _{3}}^{\left( 2n+1/2\right) }\big)
-\big( E_{-\alpha _{2}}^{\left( 2n+3/2\right) }-E_{-\alpha
_{1}-\alpha _{2}-\alpha
_{3}}^{\left( 2n+3/2\right) }\big),\vspace{2mm}  \\
G_{3}^{\left( 2n+3/2\right) }=\big( E_{\alpha _{1}+\alpha
_{2}}^{\left( 2n+1/2\right) }-E_{\alpha _{2}+\alpha _{3}}^{\left(
2n+5/2\right) }\big) -\big( E_{\alpha _{2}}^{\left( 2n+3/2\right)
}-E_{\alpha _{1}+\alpha
_{2}+\alpha _{3}}^{\left( 2n+3/2\right) }\big) +{}\vspace{1mm} \\
{}+\big( E_{-\alpha _{1}-\alpha _{2}}^{\left( 2n+5/2\right)
}-E_{-\alpha _{2}-\alpha _{3}}^{\left( 2n+1/2\right) }\big)
-\big( E_{-\alpha _{2}}^{\left( 2n+3/2\right) }-E_{-\alpha
_{1}-\alpha _{2}-\alpha
_{3}}^{\left( 2n+3/2\right) }\big),\vspace{2mm}  \\
G_{4}^{\left( 2n+1/2\right) }=-\big( E_{\alpha _{1}+\alpha
_{2}}^{\left( 2n-1/2\right) }-E_{\alpha _{2}+\alpha _{3}}^{\left(
2n+1/2\right) }\big) +\big( E_{\alpha _{2}}^{\left( 2n+1/2\right)
}-E_{\alpha _{1}+\alpha
_{2}+\alpha _{3}}^{\left( 2n+1/2\right) }\big) +{}\vspace{1mm} \\
{}+\big( E_{-\alpha _{1}-\alpha _{2}}^{\left( 2n+3/2\right)
}-E_{-\alpha _{2}-\alpha _{3}}^{\left( 2n-1/2\right) }\big)
-\big( E_{-\alpha _{2}}^{\left( 2n+1/2\right) }-E_{-\alpha
_{1}-\alpha _{2}-\alpha _{3}}^{\left( 2n+1/2\right) }\big)
\end{array}%
\right\}.
\end{gather*}

\section[Relativistic equations for $t_{\pm 3}$]{Relativistic equations for $\boldsymbol{t_{\pm 3}}$}\label{appendixB}

The decomposition of the zero curvature equation $\left[
L_{+3},L_{-3}\right] =0$ on the dif\/ferent graded subspaces reads
\begin{gather*}
\partial _{\pm 3}D^{(\mp 5/2)}  = \pm \big[ E^{(\mp 3)},B^{\mp 1}D^{(\pm
1/2)}B^{\pm 1}\big], \\
\partial _{\pm 3}D^{(\mp 2)}  = \pm \big[ D^{(\mp 5/2)},B^{\mp 1}D^{(\pm
1/2)}B^{\pm 1}\big] \pm \big[ E^{(\mp 3)},B^{\mp 1}D^{(\pm 1)}B^{\pm 1}%
\big], \\
\partial _{\pm 3}D^{(\mp 3/2)}  = \pm \big[ D^{(\mp 2)},B^{\mp 1}D^{(\pm
1/2)}B^{\pm 1}\big] \pm \big[ D^{(\mp 5/2)},B^{\mp 1}D^{(\pm 1)}B^{\pm 1}
\big]  \\
 \phantom{\partial _{\pm 3}D^{(\mp 3/2)}  =}{}
 \pm \big[ E^{(\mp 3)},B^{\mp 1}D^{(\pm 3/2)}B^{\pm 1}\big], \\
\partial _{\pm 3}D^{(\mp 1)}  = \pm \big[ D^{(\mp 3/2)},B^{\mp 1}D^{(\pm
1/2)}B^{\pm 1}\big] \pm \big[ D^{(\mp 2)},B^{\mp 1}D^{(\pm 1)}B^{\pm 1}%
\big]   \\
\phantom{\partial _{\pm 3}D^{(\mp 1)}  =}{}
\pm \big[ D^{(\mp 5/2)},B^{\mp 1}D^{(\pm 3/2)}B^{\pm 1}\big]
\pm \big[
E^{(\mp 3)},B^{\mp 1}D^{(\pm 2)}B^{\pm 1}\big], \\
\partial _{\pm 3}D^{(\mp 1/2)}  = \pm \big[ D^{(\mp 1)},B^{\mp 1}D^{(\pm
1/2)}B^{\pm 1}\big] \pm \big[ D^{(\mp 3/2)},B^{\mp 1}D^{(\pm 1)}B^{\pm 1}%
\big]  \\
\phantom{\partial _{\pm 3}D^{(\mp 1/2)}  =}{}
\pm \big[ D^{(\mp 2)},B^{\mp 1}D^{(\pm 3/2)}B^{\pm 1}\big] \pm
\big[
D^{(\mp 5/2)},B^{\mp 1}D^{(\pm 2)}B^{\pm 1}\big]   \\
\phantom{\partial _{\pm 3}D^{(\mp 1/2)}  =}{}
\pm \big[ E^{(\mp 3)},B^{\mp 1}D^{(\pm 5/2)}B^{\pm 1}\big], \\
\partial _{-3}\left( \partial _{+3}BB^{-1}\right)  = \big[
D^{(+1/2)},BD^{(-1/2)}B^{-1}\big] +\big[ D^{(+1)},BD^{(-1)}B^{-1}\big] \\
\phantom{\partial _{-3}\left( \partial _{+3}BB^{-1}\right)  =}{}+
\big[ D^{(+3/2)},BD^{(-3/2)}B^{-1}\big]
 +\big[ D^{(+2)},BD^{(-2)}B^{-1}\big] \\
\phantom{\partial _{-3}\left( \partial _{+3}BB^{-1}\right)  =}{}
 +\big[
D^{(+5/2)},BD^{(-5/2)}B^{-1}\big] +\big[
E^{(+3)},BE^{(-3)}B^{-1}\big],
\end{gather*}%
where we have replaced the solution $D^{(0)}=-\partial _{+3}BB^{-1}$
for the degree $-3$ equation
\begin{equation}
0=\big[ D^{(0)}+\partial _{+3}BB^{-1},BE^{(+3)}B^{-1}\big].
\label{-3 equation}
\end{equation}
For the $sl(2,1)_{[1]}^{(2)}$ algebra set
\begin{alignat*}{3}
& D^{(+5/2)} = \psi _{+}^{1}G_{1}^{(+5/2)},\qquad && D^{(-5/2)}=\psi _{-}^{1}G_{2}^{(-5/2)}, & \\
& D^{(+2)} = \phi _{+}^{1}M_{1}^{(+2)}, \qquad && D^{(-2)}=\phi _{-}^{1}M_{1}^{(-2)}, &  \\
& D^{(+3/2)}  = \psi _{+}^{2}F_{1}^{(+3/2)}+\psi
_{+}^{3}G_{2}^{(+3/2)}, \qquad && D^{(-3/2)}=\psi _{-}^{2}F_{2}^{(-3/2)}+\psi
_{-}^{3}G_{1}^{(-3/2)}, &  \\
& D^{(+1)} = \phi _{+}^{2}K_{1}^{(+1)}+\phi _{+}^{3}K_{2}^{(+1)}+\phi
_{+}^{4}M_{2}^{(+1)}, \qquad && D^{(-1)}=\phi
_{-}^{2}K_{1}^{(-1)}+\phi
_{-}^{3}K_{2}^{(-1)}+\phi _{-}^{4}M_{2}^{(-1)}, &  \\
& D^{(+1/2)} = \psi _{+}^{4}F_{2}^{(+1/2)}+\psi
_{+}^{5}G_{1}^{(+1/2)}, \qquad && D^{(-1/2)}=\psi _{-}^{4}F_{1}^{(-1/2)}+\psi
_{-}^{5}G_{2}^{(-1/2)}, &  \\
& D^{(0)}  =\phi _{+}^{5}M_{1}^{(0)} \qquad &&&
\end{alignat*}
then, the above equations of motion are%
\begin{gather}
\partial _{\pm 3}\psi _{\mp }^{1}  = 2\psi _{\pm }^{4}\sinh [\phi ]\pm 2\psi
_{\pm }^{5}\cosh [\phi ],  \notag \\
\partial _{\pm 3}\phi _{\mp }^{1}  = 2\psi _{\mp }^{1}\psi _{\pm }^{4}\cosh
[\phi ]\pm 2\psi _{\mp }^{1}\psi _{\pm }^{5}\sinh [\phi ]-2\phi
_{\pm
}^{3}\sinh [2\phi ]\pm 2\phi _{\pm }^{4}\cosh [2\phi ],  \notag \\
\partial _{\pm 3}\psi _{\mp }^{2}  = -\phi _{\mp }^{1}\psi _{\pm }^{4}\sinh
[\phi ]\mp \phi _{\mp }^{1}\psi _{\pm }^{5}\cosh [\phi ]+\psi _{\mp
}^{1}\phi _{\pm }^{3}\sinh [2\phi ]\mp \psi _{\mp }^{1}\phi _{\pm
}^{4}\cosh
[2\phi ],  \notag \\
\partial _{\pm 3}\psi _{\mp }^{3}  = \mp \left( \phi _{\mp }^{1}\psi _{\pm
}^{4}-2\psi _{\pm }^{3}\right) \cosh [\phi ]-\left( \phi _{\mp
}^{1}\psi _{\pm }^{5}-2\psi _{\pm }^{2}\right) \sinh [\phi ]\pm \phi
_{\pm }^{3}\psi
_{\mp }^{1}\cosh [2\phi ]  \notag \\
\phantom{\partial _{\pm 3}\psi _{\mp }^{3}  =}{}
 -\psi _{\mp }^{1}\phi _{\pm }^{4}\sinh [2\phi ]\pm \psi _{\mp
}^{1}\phi
_{\pm }^{2},  \notag \\
\partial _{\pm 3}\phi _{\mp }^{2}  = 2\left( -\psi _{\mp }^{2}\psi _{\pm
}^{4}+\psi _{\mp }^{3}\psi _{\pm }^{5}-\psi _{\mp }^{1}\psi _{\pm
}^{3}\right) \cosh [\phi ]\pm 2\left( -\psi _{\mp }^{2}\psi _{\pm
}^{5}+\psi _{\mp }^{3}\psi _{\pm }^{4}-\psi _{\mp }^{1}\psi _{\pm
}^{2}\right) \sinh
[\phi ],  \notag \\
\partial _{\pm 3}\phi _{\mp }^{3}  = 2\left( -\psi _{\mp }^{2}\psi _{\pm
}^{4}-\psi _{\mp }^{3}\psi _{\pm }^{5}+\psi _{\mp }^{1}\psi _{\pm
}^{3}\right) \cosh [\phi ]\mp 2\left( \psi _{\mp }^{2}\psi _{\pm
}^{5}+\psi _{\mp }^{3}\psi _{\pm }^{4}-\psi _{\mp }^{1}\psi _{\pm
}^{2}\right) \sinh
[\phi ]  \notag \\
  \phantom{\partial _{\pm 3}\phi _{\mp }^{3}  =}{}
  -2\phi _{\mp }^{1}\phi _{\pm }^{3}\sinh [2\phi ]\pm 2\phi _{\mp
}^{1}\phi
_{\pm }^{4}\cosh [2\phi ],  \notag \\
\partial _{\pm 3}\phi _{\mp }^{4}  = \pm 2\left( \psi _{\mp }^{2}\psi _{\pm
}^{4}+\psi _{\mp }^{3}\psi _{\pm }^{5}-\psi _{\mp }^{1}\psi _{\pm
}^{3}\right) \sinh [\phi ]+2\left( \psi _{\mp }^{2}\psi _{\pm
}^{5}+\psi _{\mp }^{3}\psi _{\pm }^{4}-\psi _{\mp }^{1}\psi _{\pm
}^{2}\right) \cosh
[\phi ]  \notag \\
\phantom{\partial _{\pm 3}\phi _{\mp }^{4}  =}{}
\pm 2\phi _{\mp }^{1}\phi _{\pm }^{3}\cosh [2\phi ]-2\phi _{\mp
}^{1}\phi
_{\pm }^{4}\sinh [2\phi ]\pm 2\phi _{\pm }^{1},  \notag \\
\partial _{\pm 3}\psi _{\mp }^{4}  = \mp \left( \phi _{\mp }^{2}\psi _{\pm
}^{4}-\phi _{\mp }^{3}\psi _{\pm }^{4}-\phi _{\mp }^{4}\psi _{\pm
}^{5}+\phi _{\mp }^{1}\psi _{\pm }^{3}\right) \cosh [\phi ]\notag\\
\phantom{\partial _{\pm 3}\psi _{\mp }^{4}  =}{}
+\left(
\phi _{\mp }^{4}\psi _{\pm }^{4}-\phi _{\mp }^{2}\psi _{\pm
}^{5}+\phi _{\mp }^{3}\psi _{\pm
}^{5}-\phi _{\mp }^{1}\psi _{\pm }^{2}\right) \sinh [\phi ]  \notag \\
\phantom{\partial _{\pm 3}\psi _{\mp }^{4}  =}{}
  \mp \left( \phi _{\pm }^{3}\psi _{\mp }^{2}+\phi _{\pm }^{4}\psi
_{\mp }^{3}\right) \cosh [2\phi ]+\left( \psi _{\mp }^{2}\phi _{\pm
}^{4}+\psi _{\mp }^{3}\phi _{\pm }^{3}\right) \sinh [2\phi ]\pm \psi
_{\mp }^{1}\phi
_{\pm }^{1}\pm \psi _{\mp }^{2}\phi _{\pm }^{2},  \notag \\
\partial _{\pm 3}\psi _{\mp }^{5}  = \mp \left( \phi _{\mp }^{4}\psi
_{+}^{4}+\phi _{\mp }^{2}\psi _{\pm }^{5}+\phi _{\mp }^{3}\psi _{\pm
}^{5}+\phi _{\mp }^{1}\psi _{\pm }^{2}-2\psi _{\pm }^{1}\right)
\cosh [\phi
]   \notag \\
\phantom{\partial _{\pm 3}\psi _{\mp }^{5}  =}{}
 -\left( \phi _{\mp }^{2}\psi _{\pm }^{4}+\phi _{\mp }^{3}\psi
_{\pm }^{4}+\phi _{\mp }^{4}\psi _{\pm }^{5}+\phi _{\mp }^{1}\psi
_{\pm }^{3}\right) \sinh [\phi ]-\left( \psi _{\mp }^{2}\phi _{\pm
}^{3}+\psi
_{\mp }^{3}\phi _{\pm }^{4}\right) \sinh [2\phi ]   \notag \\
\phantom{\partial _{\pm 3}\psi _{\mp }^{5}  =}{}
  \pm \left( \psi _{\mp }^{2}\phi _{\pm }^{4}+\psi _{\mp }^{3}\phi
_{\pm
}^{3}\right) \cosh [2\phi ]\pm \psi _{\mp }^{3}\phi _{\pm }^{2},  \notag \\
\partial _{+3}\partial _{-3}\phi  = 2\left( \psi _{+}^{4}\psi _{-}^{4}-\psi
_{+}^{5}\psi _{-}^{5}-\psi _{+}^{2}\psi _{-}^{2}+\psi _{+}^{3}\psi
_{-}^{3}-\psi _{+}^{1}\psi _{-}^{1}\right) \sinh [\phi ]   \notag \\
\phantom{\partial _{+3}\partial _{-3}\phi  =}{}
  +2\left( \psi _{+}^{4}\psi _{-}^{5}+\psi _{-}^{4}\psi
_{+}^{5}-\psi _{+}^{2}\psi _{-}^{3}-\psi _{-}^{2}\psi
_{+}^{3}\right) \cosh [\phi ]\notag \\
\phantom{\partial _{+3}\partial _{-3}\phi  =}{}
-2\left( 1-\phi _{+}^{3}\phi
_{-}^{3}+\phi _{+}^{4}\phi _{-}^{4}\right)
\sinh [2\phi ]
  -2\left( \phi _{+}^{3}\phi _{-}^{4}-\phi _{-}^{3}\phi
_{+}^{4}\right) \cosh [2\phi ]. \label{higher eq of motion}
\end{gather}
The generalized equations of motion are found by replacing above the
following solutions%
\begin{gather*}
\psi _{\pm }^{1}  = \mp 2\psi _{\pm },\qquad \psi
_{\pm }^{2}=\mp \psi _{\pm }\partial _{\pm }\phi,\qquad \psi
_{\pm
}^{3}=\partial _{\pm }\psi _{\pm },\qquad \psi _{\pm }^{4}  =  -
\frac{1}{2}\left( \psi _{\pm }\partial _{\pm }^{2}\phi -\partial
_{\pm }\phi
\partial _{\pm }\psi _{\pm }\right), \\
\psi _{\pm }^{5}  = \mp \frac{1}{2}\partial _{\pm }^{2}\psi _{\pm
}\pm \psi _{\pm }\left( \partial _{\pm }\phi \right) ^{2},\qquad
\phi _{\pm }^{1}=-\partial _{\pm }\phi,\qquad \phi _{\pm
}^{2}=-2\psi _{\pm
}\partial _{\pm }\psi _{\pm },\\
 \phi _{\pm }^{3}=\frac{1}{2}%
\left( \partial _{\pm }\phi \right) ^{2}+2\psi _{\pm }\partial _{\pm
}\psi
_{\pm }, \qquad
\phi _{\pm }^{4}  = \pm \frac{1}{2}\partial _{\pm }^{2}\phi,\\
\phi _{+}^{5}=-\frac{1}{4}\partial _{+}^{3}\phi
+\frac{1}{2}\left(
\partial _{+}\phi \right) ^{3}+3\partial _{+}\phi \psi _{+}\partial _{+}\psi
_{+}.
\end{gather*}

\subsection*{Acknowledgements}

The author thanks FAPESP and CNPq for partial f\/inancial support, J.F.~Gomes for comments and the referees for useful suggestions.
The author also thanks Alexis Roa and Suzana Moreira for reading the
manuscript.

\pdfbookmark[1]{References}{ref}
\LastPageEnding

\end{document}